\documentclass[useAMS,usegraphicx,usenatbib]{mn2e}

\usepackage{subfig}

\newcommand{\de}{$\degr$}
\newcommand{\kms}{\hbox{km s$^{-1}$}}
\newcommand{\vsini}{\hbox{$v\,\sin\,i$}}

\title[Doppler images of ER Vul]{Doppler images of the eclipsing binary ER Vulpeculae}
\author[Y.~Xiang, S.~Gu, A.~Collier~Cameron and J.~R.~Barnes]{Yue Xiang,$^{1}$$^{2}$$^{3}$\thanks{E-mail: xy@ynao.ac.cn} Shenghong Gu,$^{1}$$^{2}$ A.~Collier~Cameron$^{4}$ and J.~R.~Barnes$^{5}$\\
$^{1}$Yunnan Observatories, Chinese Academy of Sciences, Kunming 650011, China\\
$^{2}$Key Laboratory for the Structure and Evolution of Celestial Objects, Chinese Academy of Sciences, Kunming 650011, China\\
$^{3}$University of Chinese Academy of Sciences, Beijing 100049, China\\
$^{4}$School of Physics and Astronomy, University of St Andrews, Fife KY16 9SS, UK\\
$^{5}$Center for Astrophysics Research, University of Hertfordshire, College Lane, Hatfield, Hertfordshire AL10 9AB, UK}

\begin{document}


\maketitle

\label{firstpage}

\begin{abstract}
We present Doppler images of both components of the eclipsing binary system ER Vul, based on the spectra obtained in 2004 November, 2006 September and 2008 November. The least-squares deconvolution technique is used for enhancing the signal-to-noise ratios of the observed profiles. The new surface images reveal that both stars of ER Vul show strong starspot activities and the starspots appear at various latitudes. The surface maps of 2006 and 2008 both show the presence of large high-latitude starspots on each component of ER Vul. We find no obvious phase shift of the active regions during our observations. The longitude distributions of starspots are non-uniform on both stars. At low-to-mid latitudes, the active regions are almost exclusively found in the hemisphere facing the other star. However, we find no pronounced concentration of spots at the sub-stellar points.
\end{abstract}


\section{Introduction}

ER Vul is an active eclipsing binary which is composed of two Sun-like stars, G0V primary and G5V secondary, in an orbital period of 0.698 d. According to \citet{har2004}, the masses of the primary and the secondary stars are 1.02 M$_{\odot}$ and 0.97 M$_{\odot}$, respectively. \citet{hall1976} classified ER Vul as a short-period RS CVn-type binary system. \citet{kju2003} inferred that both components of ER Vul have filled half of their Roche lobes. However, \citet{due2003} suggested that the primary star nearly fills its Roche lobe, but the binary system is still detached. \citet*{dry2005} identified ER Vul as a pre-contact binary system, which belongs to the short-period RS CVn binaries and has similar property to W UMa-type contact binaries.

As a member of RS CVn binaries, ER Vul shows strong chromospheric and starspot activities. From several chromospheric activity indicators, \citet{laz1997} revealed that both component stars of ER Vul are active. \citet{cak2003} analyzed the contribution of H$_{\alpha}$ emission from each component, and found that the secondary star is more active than the primary one. ER Vul also shows X-ray, EUV and radio emissions \citep*{ost2002,san2003}, which reveal continuous activities, such as microflares in the system \citep{gunn1997}. Starspots were found on both components of ER Vul by using light curve modelling \citep*{olah1994,ekm2002,wil2011}. \citet{pop2013} revealed an orbital period change of ER Vul from the photometric data spanning 54.9 yr. They inferred the period change is due to the cyclic magnetic activity on both components or the third body in the system.

ER Vul is a suitable target for Doppler imaging owing to the high rotational velocity. \citet{pis1996} and \citet{pis2001} presented the Doppler images of ER Vul based on the spectra obtained in 1993, 1994 and 1996 seasons. Their surface maps showed large temperature variation on both component stars of ER Vul. They also found hot spots at the sub-stellar points on both stars, and suggested that the presence of these bright features is due to the reflecting effect. \citet{pis2008} analyzed the persistent cool starspots in Doppler images and revealed a non-axisymmetric dynamo action in ER Vul, which may be due to the tidal interaction.

We have carried out a series of high-resolution spectroscopic observations for various binary systems to study the starspot activities of stars with different parameters and evolutionary stages. We had presented some results on the single-lined RS CVn-type binary II Peg \citep{gu2003,xiang2014}. In order to study the starspot activities on both stars of a close binary system and how the activities are affected by the interaction between two components, we choose the double-lined binary ER Vul as the target star for Doppler imaging in this work. Both components of ER Vul are imaged simultaneously based on the spectroscopic data obtained in 2004, 2006 and 2008. We shall describe the observation and data reduction in Section 2. The results will be given and discussed in Section 3 and 4, respectively. In Section 5, we shall summarize the present work.

\section{Observations and data reduction}

The new spectroscopic observations were carried out on 2004 November 20-29, 2006 September 1-7 and 2008 November 13-17, using the Coud\'{e} echelle spectrograph \citep{zhao2001} of the 2.16m telescope at the Xinglong station of National Astronomical Observatories, China. A 1024$\times$1024 pixel TEK CCD was equipped in the spectrograph to record data. The spectrum covers the wavelength range from about 5622\AA\ to 9000\AA\, and has a resolution of R = 37000. Most observations in 2006 and 2008 had exposure times of 900--1200s, which correspond to 1.5\%--2.0\% of the orbital period, but the observations in 2004 had exposure times of 2400--3600s to get enough S/N. We summary the detailed observational log, including the heliocentric Julian dates, phases and exposure times in Table \ref{tab:log}. The orbital phases were calculated based on the ephemeris,

T = HJD 2445221.10924 + 0.698094903E, \\derived by \citet{pop2013}. The primary star is behind the secondary one at $T_{0}$. Besides, a small phase offset for each dataset was also considered as described in Section 3.1.

\begin{table}
 \caption{The observing log.}
 \label{tab:log}
 \begin{tabular}{lccccc}
  \hline
  Date   & HJD  & Phase & Exp. & S/N & S/N\\
  &2450000+&&(s)&Input&LSD\\
  \hline
  20/11/2004 & 3330.0088 & 0.7590 & 2400 & 195 & 2269\\
  21/11/2004 & 3331.0117 & 0.1957 & 2400 & 165 & 1872\\
  27/11/2004 & 3336.9727 & 0.7346 & 3600 & 231 & 2653\\
  28/11/2004 & 3337.9739 & 0.1688 & 3000 & 147 & 1660\\
  29/11/2004 & 3338.9814 & 0.6120 & 3000 & 112 & 1266\\

  01/09/2006 & 3980.0907 & 0.9781 & 1200 & 121 & 1421 \rule{0pt}{12pt}\\
  01/09/2006 & 3980.1050 & 0.9986 & 1200 & 127 & 1488\\
  01/09/2006 & 3980.1192 & 0.0190 & 1200 & 116 & 1346\\
  01/09/2006 & 3980.1336 & 0.0395 & 1200 & 102 & 1176\\
  04/09/2006 & 3983.0665 & 0.2409 & 1200 &  43 &  530\\
  04/09/2006 & 3983.0808 & 0.2613 & 1200 &  61 &  754\\
  04/09/2006 & 3983.0953 & 0.2821 & 1200 &  77 &  936\\
  04/09/2006 & 3983.1110 & 0.3047 & 1200 &  66 &  809\\
  05/09/2006 & 3984.0632 & 0.6687 & 1200 & 115 & 1403\\
  05/09/2006 & 3984.0773 & 0.6888 & 1200 & 123 & 1501\\
  05/09/2006 & 3984.0914 & 0.7090 & 1200 & 127 & 1547\\
  05/09/2006 & 3984.1054 & 0.7292 & 1200 & 126 & 1544\\
  06/09/2006 & 3985.0582 & 0.0939 & 1200 & 136 & 1697\\
  06/09/2006 & 3985.0723 & 0.1141 & 1200 & 140 & 1734\\
  06/09/2006 & 3985.0864 & 0.1343 & 1200 & 146 & 1807\\
  06/09/2006 & 3985.1007 & 0.1548 & 1200 & 148 & 1834\\
  06/09/2006 & 3985.1634 & 0.2446 & 1200 & 148 & 1849\\
  06/09/2006 & 3985.1775 & 0.2648 & 1200 & 146 & 1819\\
  06/09/2006 & 3985.2468 & 0.3641 & 1200 & 146 & 1807\\
  06/09/2006 & 3985.2608 & 0.3842 & 1200 & 142 & 1741\\
  07/09/2006 & 3986.0885 & 0.5698 & 1200 &  69 &  822\\
  07/09/2006 & 3986.0997 & 0.5859 &  717 &  55 &  660\\

  13/11/2008 & 4783.9710 & 0.5115 &  900 & 121 & 1491 \rule{0pt}{12pt}\\
  13/11/2008 & 4783.9821 & 0.5274 &  900 & 118 & 1461\\
  13/11/2008 & 4784.0728 & 0.6573 &  900 &  96 & 1159\\
  13/11/2008 & 4784.0839 & 0.6731 &  900 &  98 & 1159\\
  13/11/2008 & 4784.1412 & 0.7552 & 1200 &  98 & 1123\\
  14/11/2008 & 4784.9444 & 0.9059 &  900 &  78 &  922\\
  14/11/2008 & 4784.9551 & 0.9211 &  900 &  76 &  894\\
  14/11/2008 & 4785.0124 & 0.0032 &  900 &  64 &  748\\
  14/11/2008 & 4785.0234 & 0.0189 &  900 &  73 &  853\\
  14/11/2008 & 4785.0839 & 0.1057 &  900 &  70 &  797\\
  14/11/2008 & 4785.0948 & 0.1212 &  900 &  66 &  735\\
  14/11/2008 & 4785.1324 & 0.1751 &  900 &  57 &  641\\
  14/11/2008 & 4785.1430 & 0.1903 &  900 &  53 &  584\\
  15/11/2008 & 4785.9323 & 0.3209 &  900 &  95 & 1196\\
  15/11/2008 & 4785.9432 & 0.3365 &  900 &  91 & 1143\\
  15/11/2008 & 4786.0021 & 0.4210 &  900 &  88 & 1087\\
  15/11/2008 & 4786.0129 & 0.4365 &  900 &  81 & 1010\\
  15/11/2008 & 4786.0673 & 0.5144 &  900 &  84 & 1029\\
  15/11/2008 & 4786.0781 & 0.5298 &  900 &  85 & 1044\\
  15/11/2008 & 4786.1235 & 0.5949 &  900 &  87 & 1021\\
  15/11/2008 & 4786.1343 & 0.6103 &  900 &  83 &  968\\
  16/11/2008 & 4786.9364 & 0.7593 &  900 & 101 & 1242\\
  16/11/2008 & 4786.9472 & 0.7748 &  900 &  95 & 1168\\
  16/11/2008 & 4787.0070 & 0.8604 &  900 &  93 & 1153\\
  16/11/2008 & 4787.0182 & 0.8764 &  900 &  90 & 1111\\
  16/11/2008 & 4787.0841 & 0.9709 &  900 &  53 &  641\\
  16/11/2008 & 4787.0948 & 0.9862 &  900 &  52 &  624\\
  17/11/2008 & 4787.9389 & 0.1954 &  900 &  69 &  849\\
  17/11/2008 & 4787.9497 & 0.2109 &  900 &  71 &  875\\
  17/11/2008 & 4788.0098 & 0.2969 &  900 &  73 &  917\\
  17/11/2008 & 4788.0207 & 0.3125 &  900 &  73 &  889\\
  17/11/2008 & 4788.0724 & 0.3867 &  900 &  80 &  957\\
  17/11/2008 & 4788.0834 & 0.4023 &  900 &  80 &  945\\
  17/11/2008 & 4788.1249 & 0.4618 & 1200 &  80 &  932\\
  \hline
 \end{tabular}
\end{table}

Because the two-temperature model is used in our image reconstruction, we also observed several inactive slowly rotating template stars by using the same instrument setup as ER Vul. We use the spectra of HR 4277 (G0V) and HR 3309 (G5V) to represent the spectra of the photospheres of the primary star (T$_{eff}$ = 6000K) and the secondary star (T$_{eff}$ = 5750K), respectively. We set the starspot temperature to T$_{spot}$ = 5000K, based on the temperature difference between the starspot and the photosphere, derived from light curve modelling \citep{ekm2002,wil2011} and Doppler imaging \citep{pis2001}, for both component stars. We use the spectrum of HR 568 (K0IV) to represent the spectrum of the starspot.

The spectroscopic data were reduced with IRAF\footnote{IRAF is distributed by the National Optical Astronomy Observatory, which is operated by the Association of Universities for Research in Astronomy (AURA) under cooperative agreement with the National Science Foundation.} package in a standard way. The reduction procedure included image trimming, bias subtraction, flat-field dividing, scatter light subtraction, cosmic-ray removal, 1D spectrum extraction, wavelength calibration and continuum fitting. The wavelength calibration was carried out using the comparison spectra of the ThAr lamp for each night.

\begin{figure}
\centering
\includegraphics[width=0.43\textwidth]{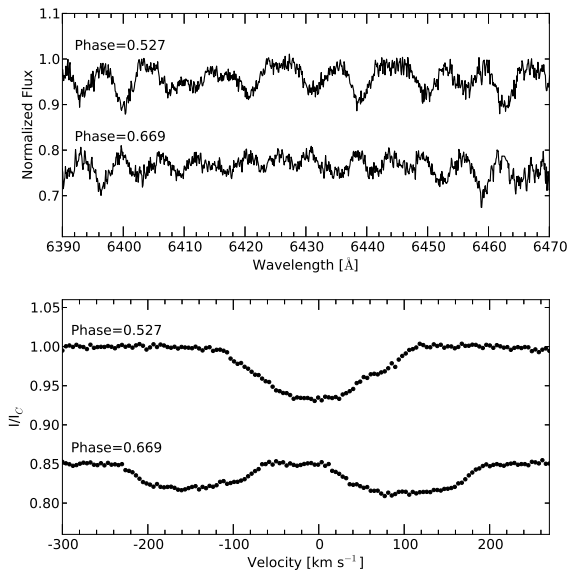}
\caption{Examples for the input spectra (upper panel) and the corresponding LSD profiles (lower panel) at orbital phases 0.527 and 0.669. In order to show clearly, the input spectra are just parts around 6400\AA.}
\label{fig:example}
\end{figure}

In order to significantly enhance the signal-to-noise ratios (S/N) of the observed spectra, we used the Least-Squares Deconvolution (LSD; \citealt{don1997}) technique to combine all available photospheric lines in each spectrum. The line list for ER Vul was obtained from Vienna Atomic Line Database (VALD; \citealt{kup1999}). Some physical parameters in Table \ref{tab:par} were used to generate the line list. The value of microturbulent velocity ($\xi$) was estimated from the values of the early G-type dwarfs in \citet{gray2001}. The wavelength regions which contain strong chromospheric and telluric lines, were excluded from the line list. During LSD calculations, we set the velocity increment per pixel to 4.1 \kms, according to the spectral resolution power. Instrument shifts in the wavelength calibration were corrected with the telluric lines, using the method described by \citet{cam1999}. All spectra in one dataset have the same radial velocity zero-point after such corrections. This method has been demonstrated to have a precision better than 0.1 \kms\ \citep{don2003}. The input peak S/N and output S/N of each spectrum are listed in Table \ref{tab:log}. As examples, we plot the input spectra and the corresponding LSD profiles in Fig. \ref{fig:example}. Besides, the spectra of the template stars were also deconvolved in the same manner to mimic the local intensity profiles of the photospheres and the spots. We used linear interpolation of the limb-darkening coefficients derived by \citet*{cla2013} for UBVRI passbands to obtain the values for the photosphere and spot temperatures of each component star, and 30 limb angles were used for producing the lookup tables.

\section{Doppler imaging}
\subsection{System parameters}

\begin{table}
 \centering
 \caption{Adopted stellar parameters of ER Vul for Doppler imaging.}
 \label{tab:par}
 \begin{tabular}{lcc}
 \hline
 Parameter & Value & Ref.\\
 \hline
 $q=M_{2}/M_{1}$ & 0.9518 & a\\
 $K_{1}$ (km s$^{-1}$) & 135.33 & a\\
 $K_{2}$ (km s$^{-1}$) & 142.18 & K$_{1}$ and $q$\\
 $\gamma$ (km s$^{-1}$) & -26.9/-26.1/-27.9 & DoTS\\
 $i$ (\de) & 66.63 & a\\
 $T_{0}$ (HJD) & 2445221.10924 & b\\
 $P$ (d)  & 0.698094903 & b\\
 \vsini~$_{1}$ (km s$^{-1}$) & 83.9 & DoTS\\
 \vsini~$_{2}$ (km s$^{-1}$) & 77.9 & DoTS\\
 $T_{eff,1}$ (K)& 6000 & a\\
 $T_{eff,2}$ (K)& 5750 & a\\
 log $g$ & 4.5 & a\\
 $\xi$ (km s$^{-1}$) & 1.5 & c\\
 $[$Fe/H$]$ & -0.29 & d\\
 albedo$_{1,2}$ & 0.5 & a\\
 \hline
 a. \citet{har2004}\\
 b. \citet{pop2013}\\
 c. \citet{gray2001}\\
 d. \citet{nor2004}\\
 \end{tabular}
\end{table}

Doppler imaging is very sensitive to the adopted values of stellar parameters, and wrong parameters will lead to artifacts in the reconstructed images \citep{cam1994}. The accurate orbital parameters are very important for imaging the eclipsing binary systems \citep*{vin1993}. It was demonstrated that the Doppler imaging code can be used for determining stellar parameters for both single and binary stars, and can avoid systematic errors caused by the effects of starspot distortions \citep{bar1998,bar2004}. In our case, we used the Doppler imaging code DoTS (Doppler Tomography of Star; \citealt{cam1997}) to perform a fixed number of iterations to get solutions, and then searched the best-fit parameters which lead to a minimal $\chi^{2}$ \citep{xiang2014}.

More parameters need to be determined for binary systems than that for single stars. Fortunately, ER Vul has been extensively studied, and the physical and orbital parameters have been derived by many authors. We adopted the improved values of the radial velocity amplitude of the primary star ($K_{1}$), mass ratio (q) and inclination (i) derived from both the photometric data and the radial velocities \citep{har2004}. The first conjunction time ($T_{0}$) and the orbital period (P) were from the paper by \citet{pop2013}. The reflection effect was also considered by the imaging code, as described by \citet{cam1997}. We adopted the values of the albedo coefficients from \citet{har2004}.

After fixing the above well-determined parameters, we used the Doppler imaging code DoTS and $\chi^{2}$ minimization method to search the best-fit values of the projected rotational velocities (\vsini) for two components and the radial velocity of mass center ($\gamma$). \vsini\ was determined by using the dataset obtained in 2008 November, which had the best phase coverage in three observing runs. The final values we adopted are 83.9 \kms\ and 77.9 \kms, which are between the ones derived by \citet*{hill1990} (81 \kms\ and 71 \kms) and \citet{kju2003} (90 \kms\ and 80 \kms). The value of $\gamma$ was determined for each dataset, because of the different instrumental radial velocity zero-points of three observing runs. The orbital period change of ER Vul can produce phase offsets in its ephemeris. The image code can be used to search the best-fit value of the phase offset \citep{hus2006,xiang2014}. The resulting values are -0.0037, 0 and 0.0011 for our three datasets, which result in offsets less than 1.5\de\ in longitude of the Doppler images. This has no significant effect on the reconstructed image. We list the adopted values of stellar parameters for Doppler imaging of ER Vul in Table \ref{tab:par}.

\begin{figure*}
\centering
\includegraphics[width=0.96\textwidth]{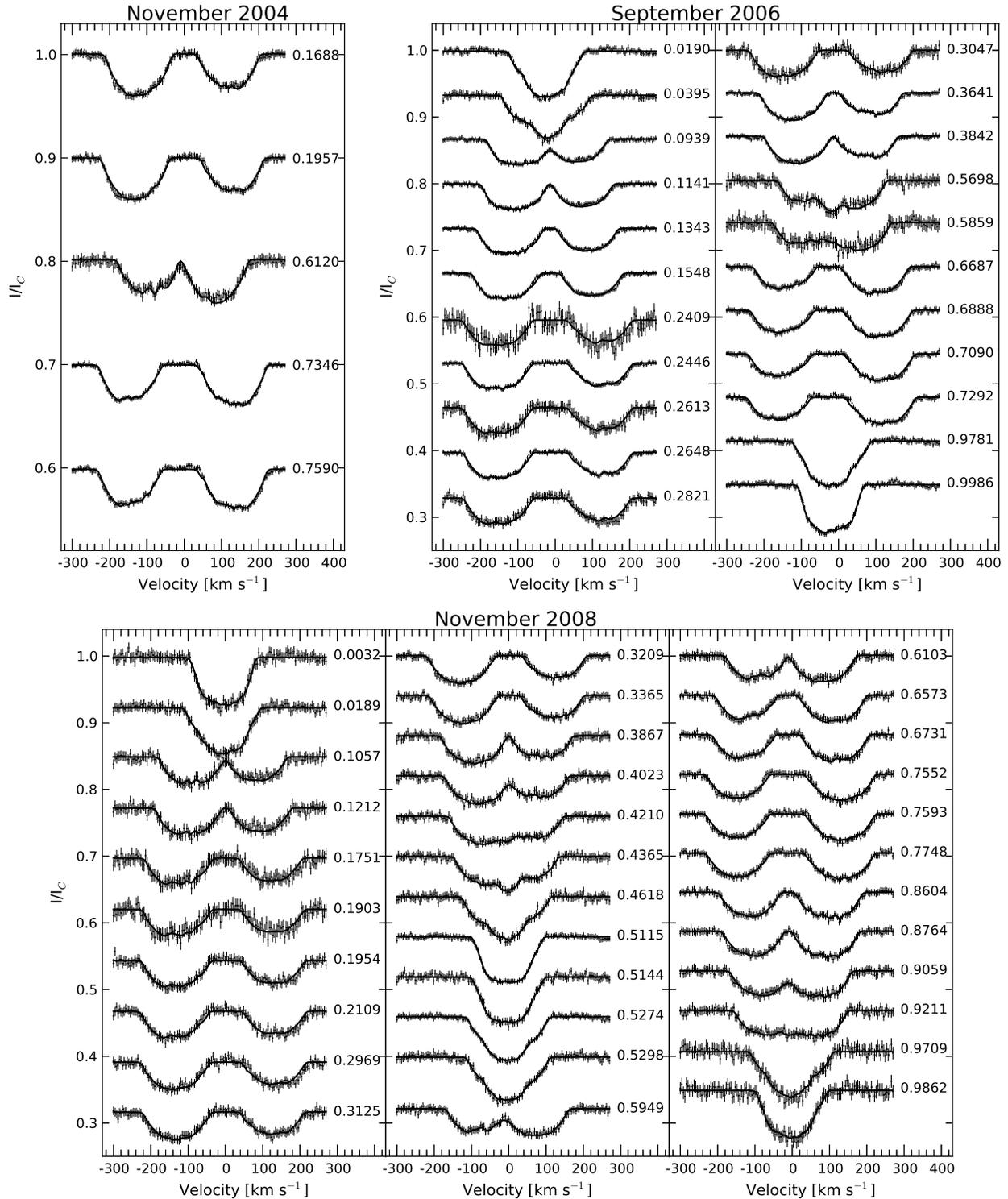}
\caption{The LSD profiles (dots with error bars) and the maximum entropy solutions (solid lines). The observing phase is also annotated beside each profile.}
\label{fig:profiles}
\end{figure*}

\subsection{Results}

\begin{figure*}
\centering
\includegraphics[angle=270,width=0.85\textwidth]{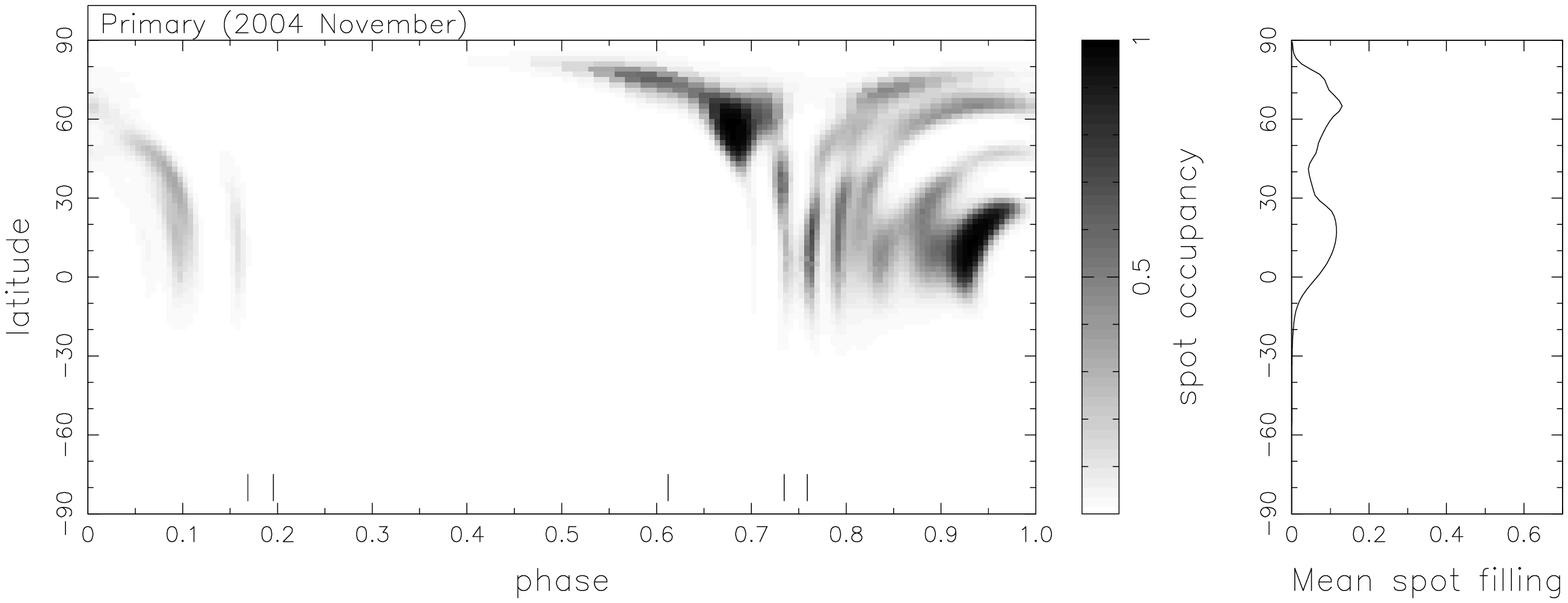}
\includegraphics[angle=270,width=0.85\textwidth]{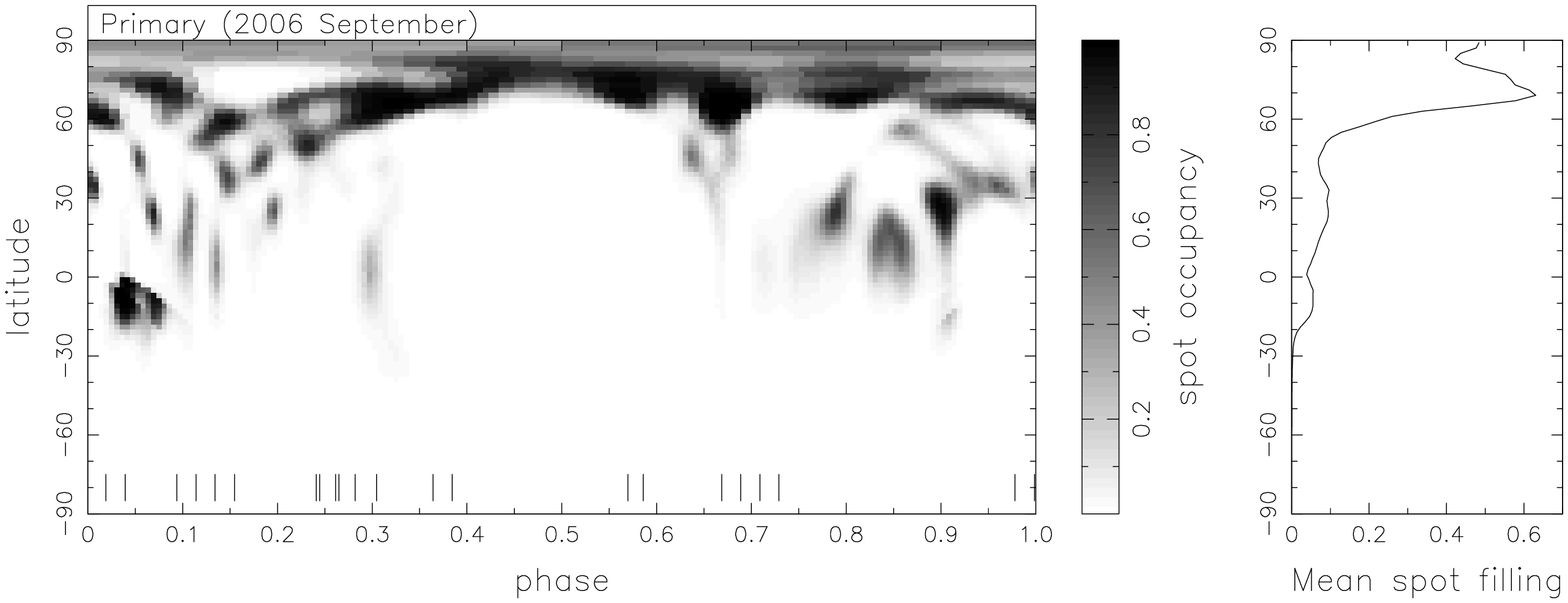}
\includegraphics[angle=270,width=0.85\textwidth]{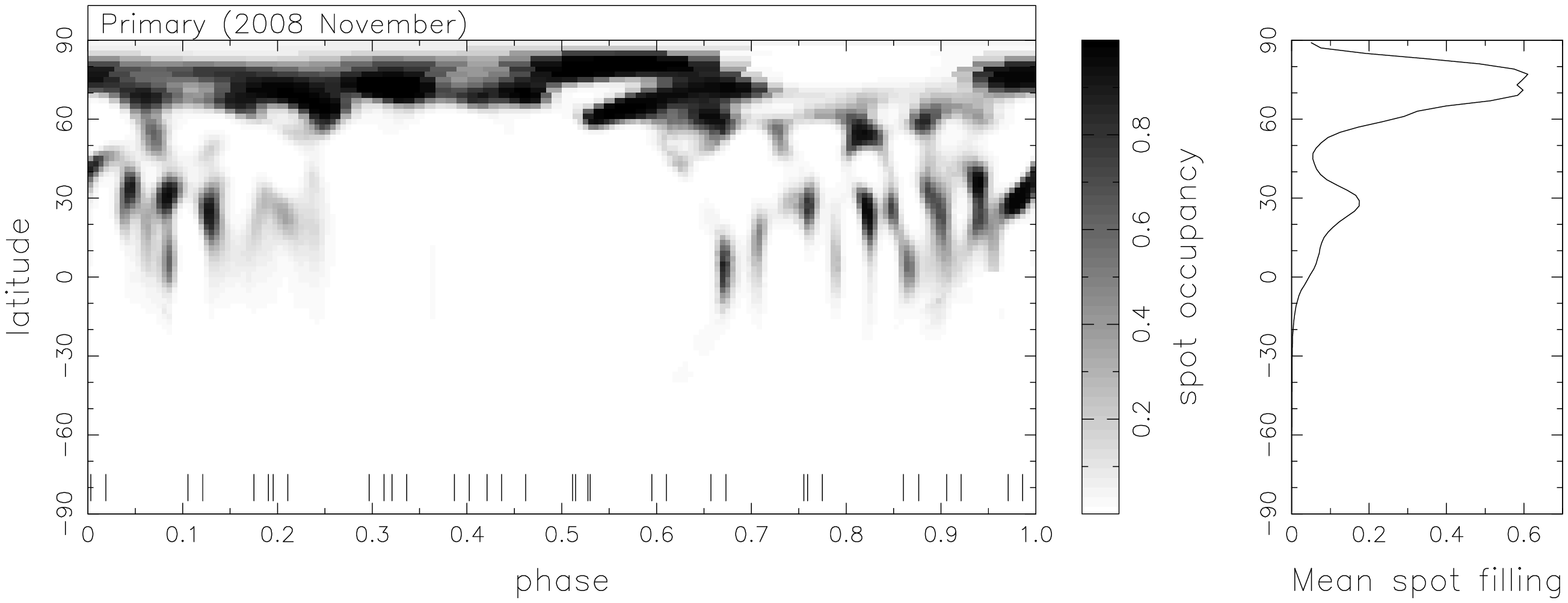}
\caption{Mercator projection of the reconstructed images for the primary star in 2004 (upper), 2006 (middle) and 2008 (bottom). Observing phases are marked as ticks. The mean spot filling factor as a function of latitude is also plotted on the right of each image.}
\label{fig:primary}
\end{figure*}

\begin{figure*}
\centering
\includegraphics[angle=270,width=0.85\textwidth]{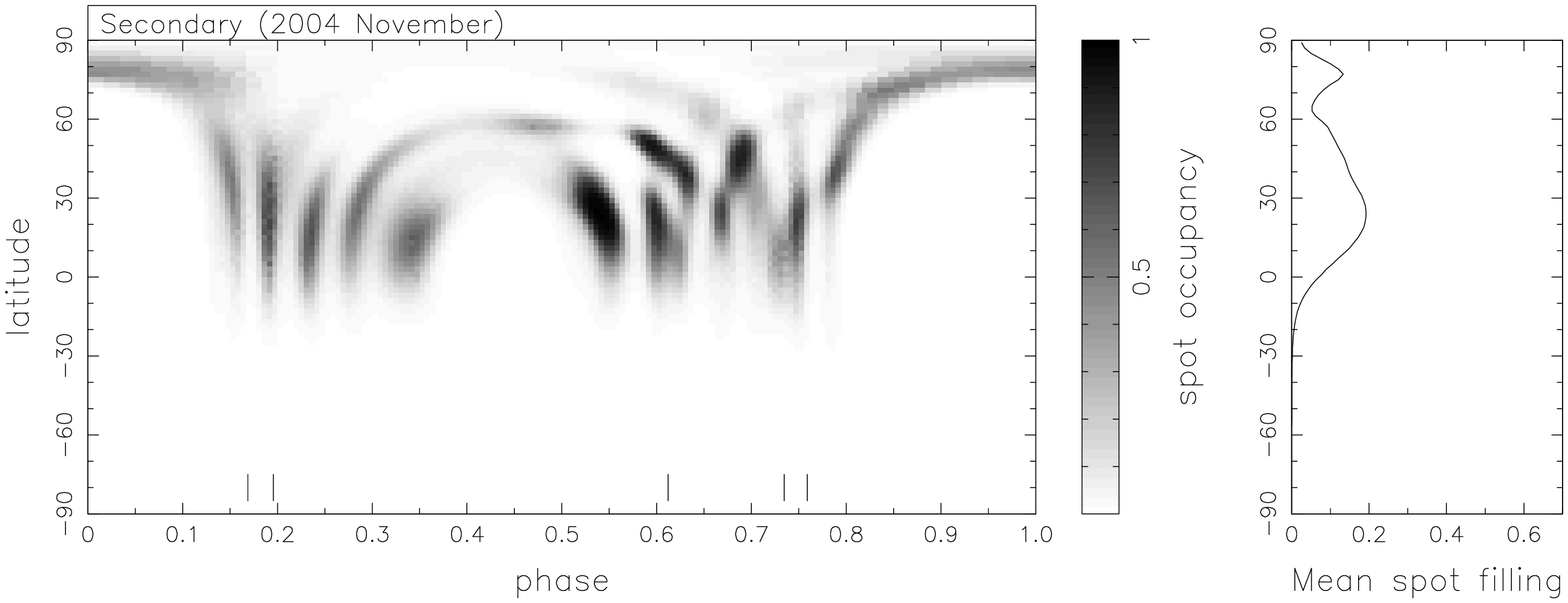}
\includegraphics[angle=270,width=0.85\textwidth]{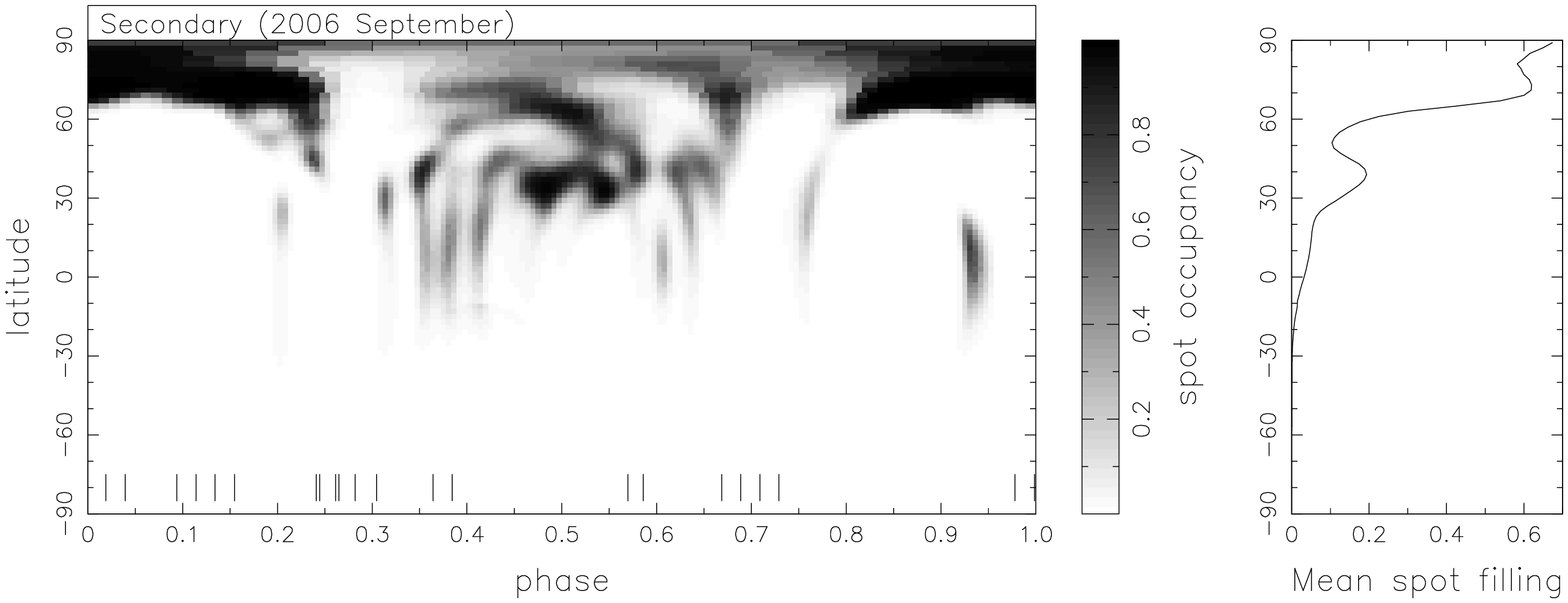}
\includegraphics[angle=270,width=0.85\textwidth]{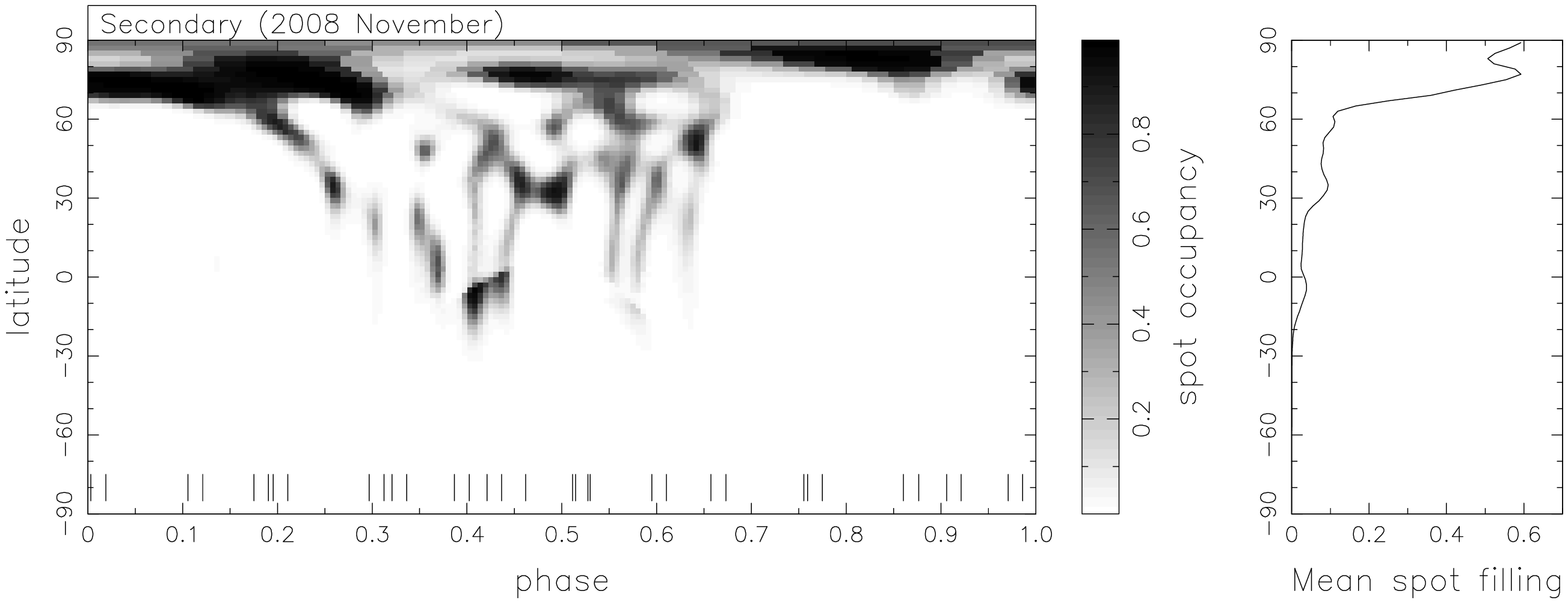}
\caption{Same as Fig. \ref{fig:primary}, but for the secondary star.}
\label{fig:secondary}
\end{figure*}

For three datasets, we used the maximum entropy Doppler imaging code DoTS to reconstruct the surface images of both components of ER Vul simultaneously. The LSD profiles and the maximum entropy solutions are both plotted in Fig. \ref{fig:profiles}. We present the Mercator projections of the Doppler images of the primary and the secondary components in Fig. \ref{fig:primary} and Fig. \ref{fig:secondary}, respectively. The mean spot filling factor as the function of latitude is also plotted beside each spot map. In our surface images, phase 0 on the primary star faces phase 0.5 on the secondary star.

In 2004 November, the primary star showed a strong high-latitude spot at phase 0.7 and low-latitude spots at phase 0.95. Some weak spot structures were located between phases 0.7 and 1. The secondary star had several strong low- and intermediate-latitude spots at phase 0.6--0.7. There were also weak spots at phase 0.2--0.4 and 0.6--0.8. However, because of the poor phase coverage, much information of starspots was lost in 2004 November. The phase coverage was good in 2006 September and 2008 November. In 2006 September, the primary star had several strong starspots at high latitudes. Some appendages extending to intermediate and low latitude regions were attached to high-latitude spots. At low latitudes, there were also some starspots at phase 0.0--0.3 and 0.7--1.0. The spot map of the primary star in 2008 is similar to the one in 2006, but there were more low-latitude spots in 2008 November. On the secondary star, a strong extended high-latitude spot between phases 0.8 and 1.2 and a smaller one at phase 0.5 could be seen in the surface map of 2006 September. There were also some low-latitude spots at phase 0.3--0.7. In 2008 November, the spot map was similar to the one in 2006, but the large high-latitude starspot seemed to become two spots, and the strength of low-latitude spots changed much.

\begin{figure*}
\centering
\subfloat[]{
\label{fig:test:all}
\begin{tabular}{cc}
\includegraphics[angle=270,width=0.45\textwidth]{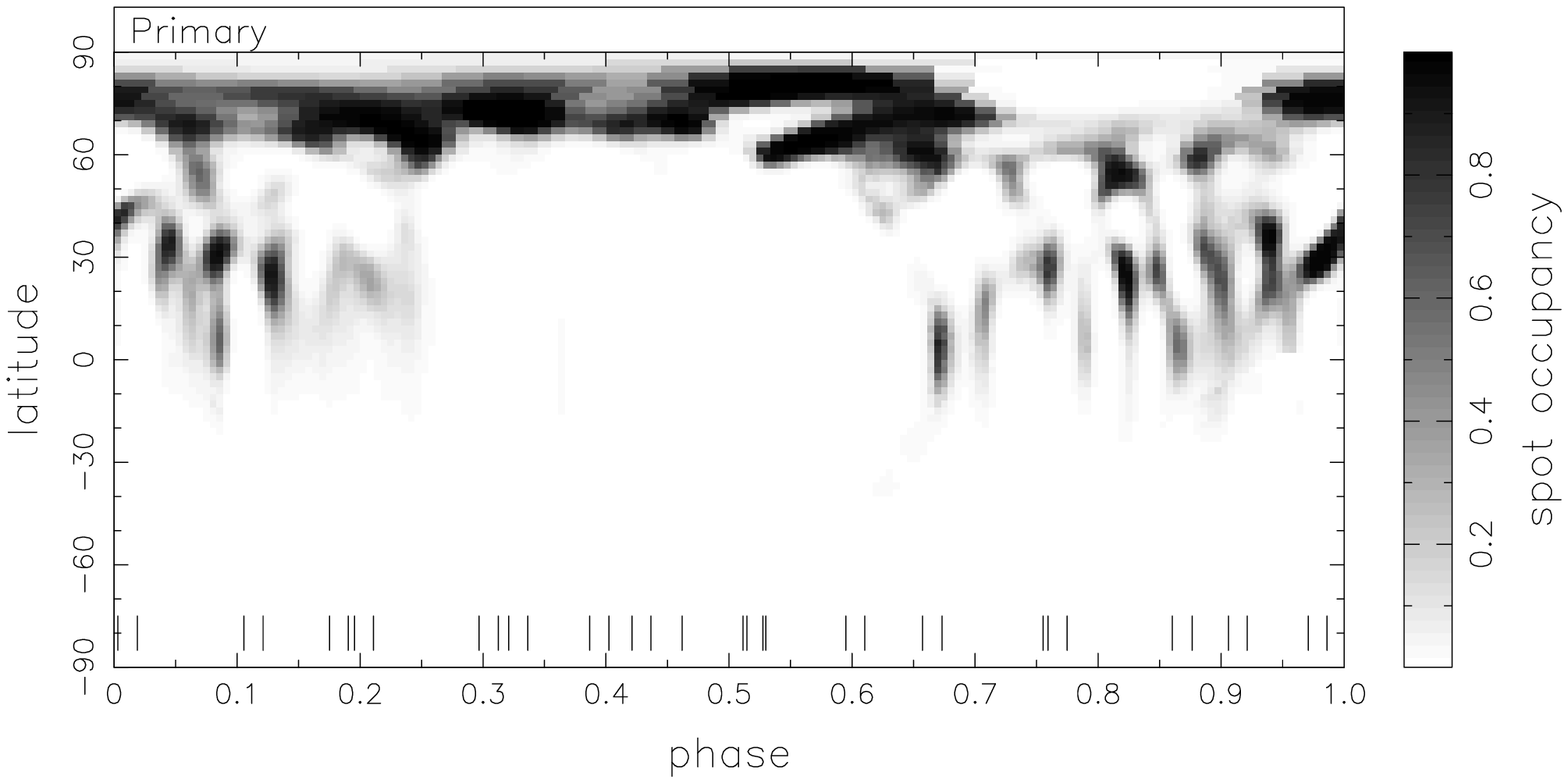}
&
\includegraphics[angle=270,width=0.45\textwidth]{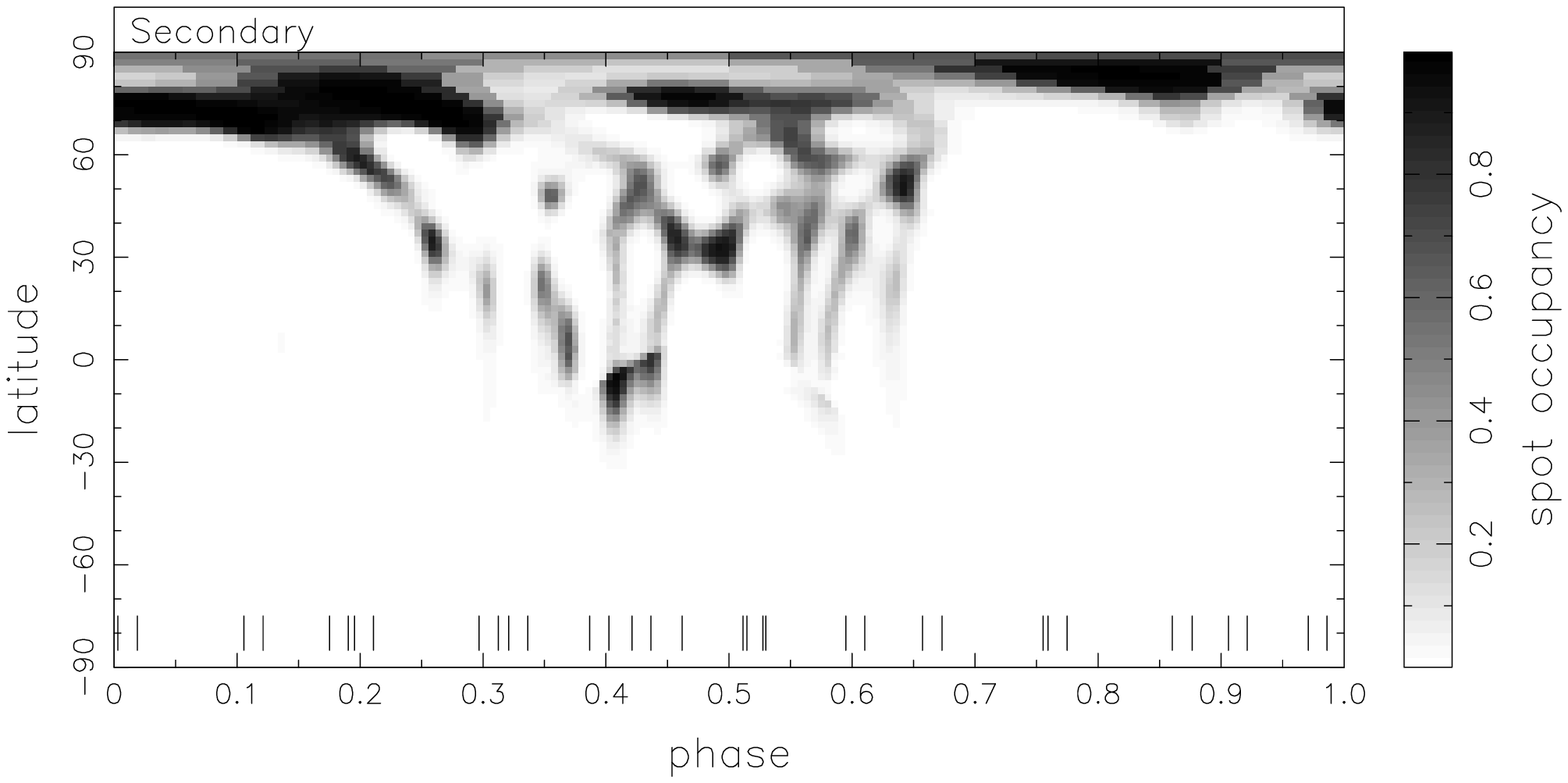}
\\
\end{tabular}
}

\subfloat[]{
\label{fig:test:odd}
\begin{tabular}{cc}
\includegraphics[angle=270,width=0.45\textwidth]{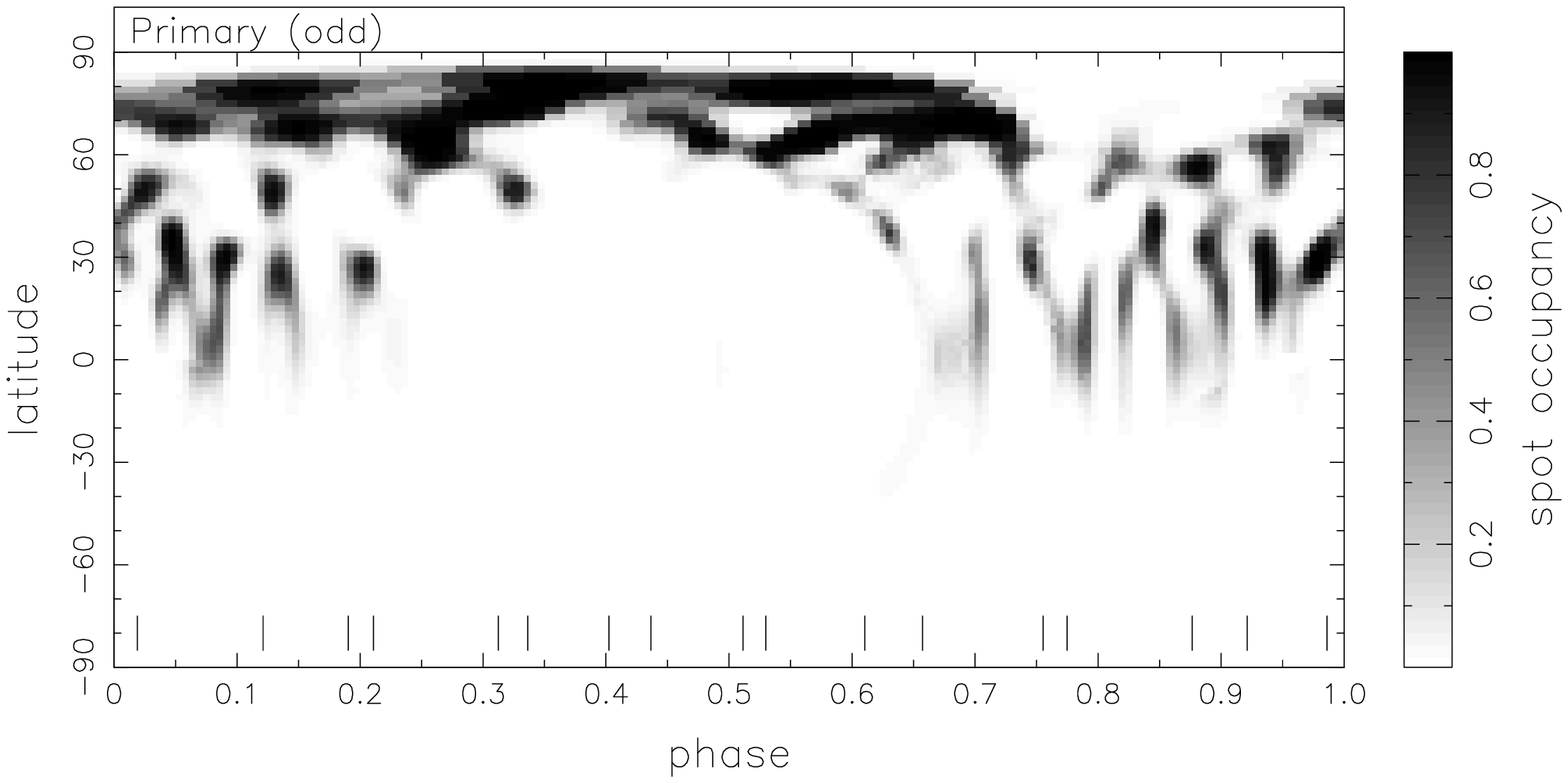}
&
\includegraphics[angle=270,width=0.45\textwidth]{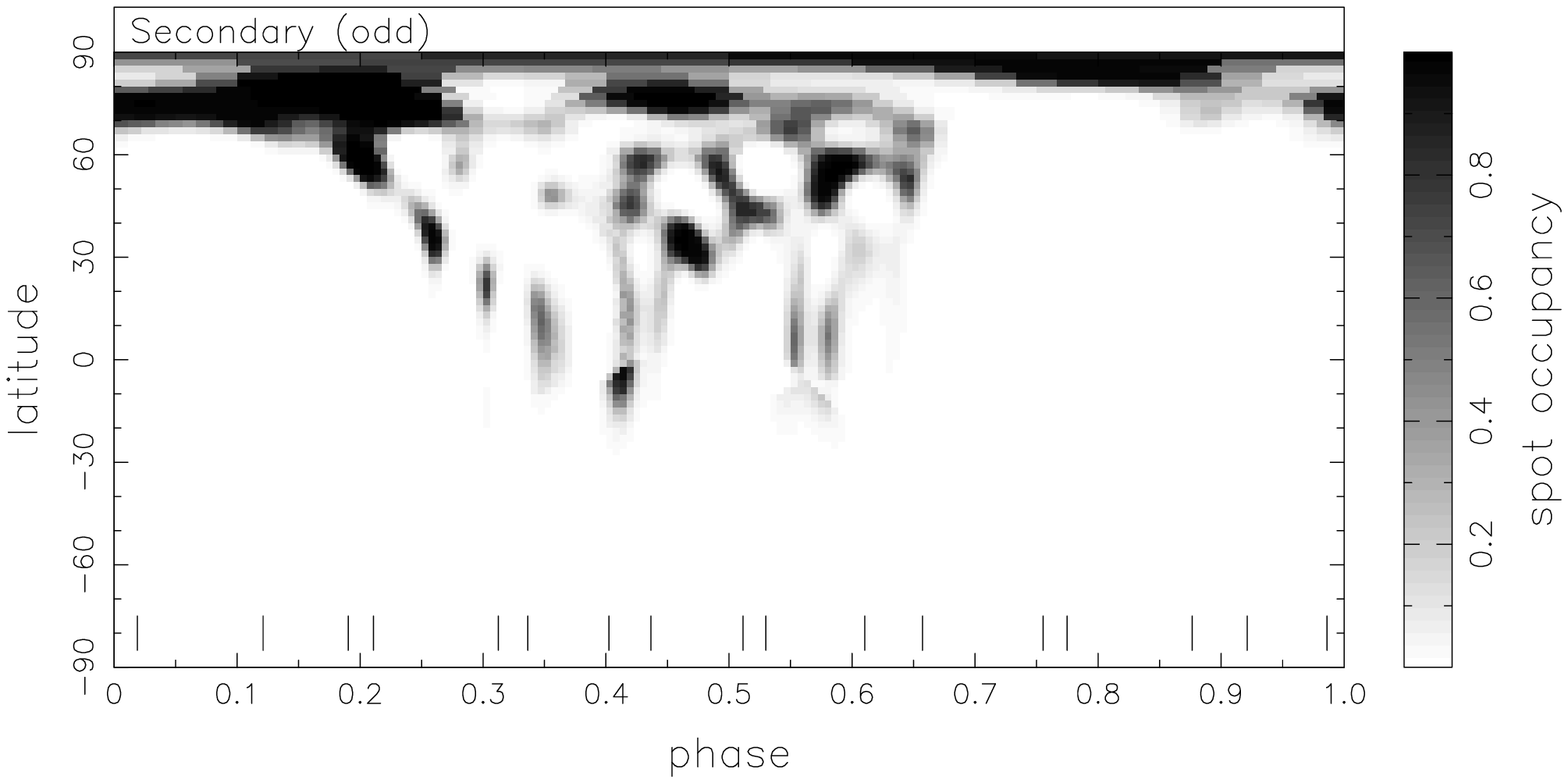}
\\
\end{tabular}
}

\subfloat[]{
\label{fig:test:even}
\begin{tabular}{cc}
\includegraphics[angle=270,width=0.45\textwidth]{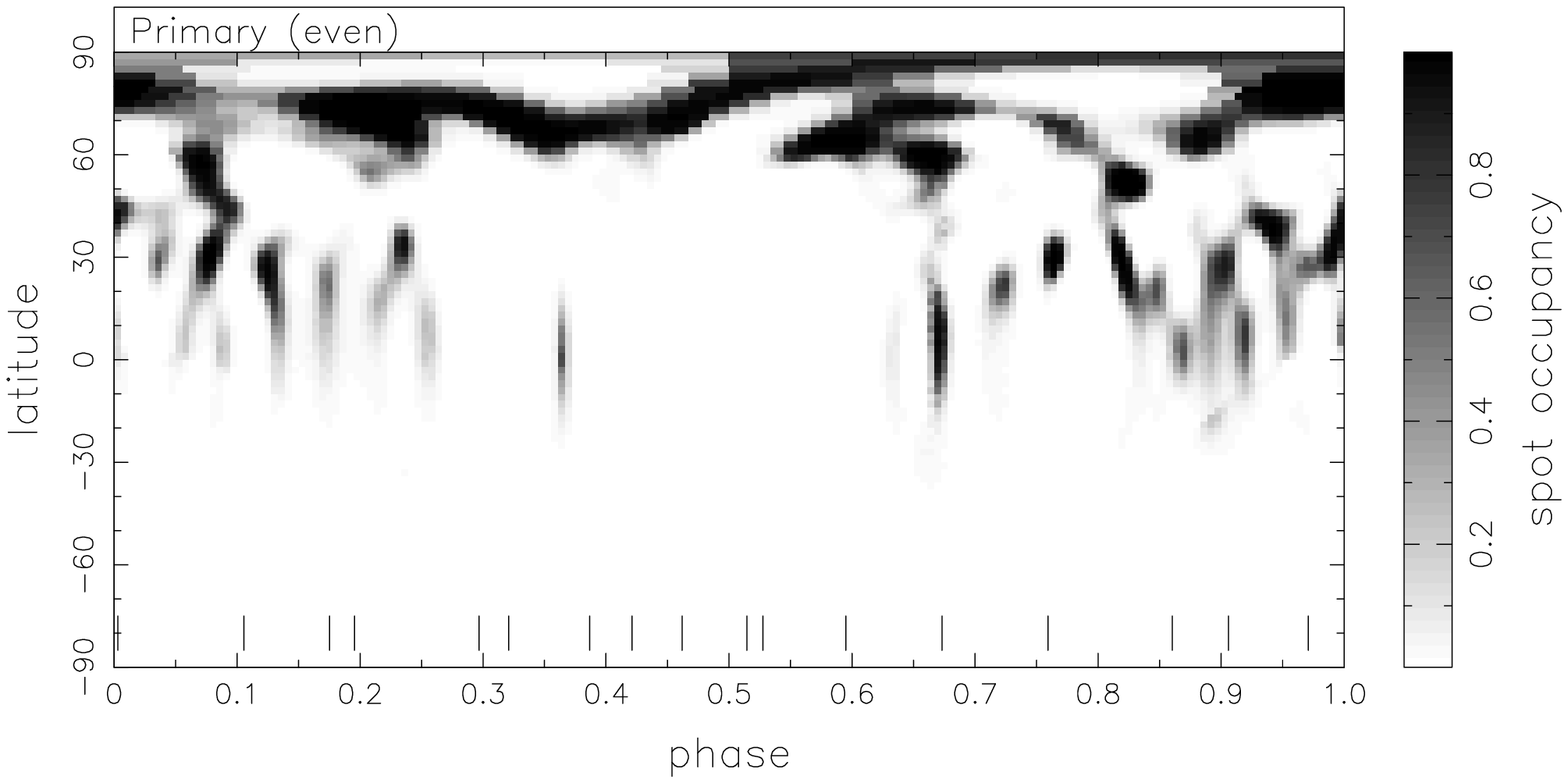}
&
\includegraphics[angle=270,width=0.45\textwidth]{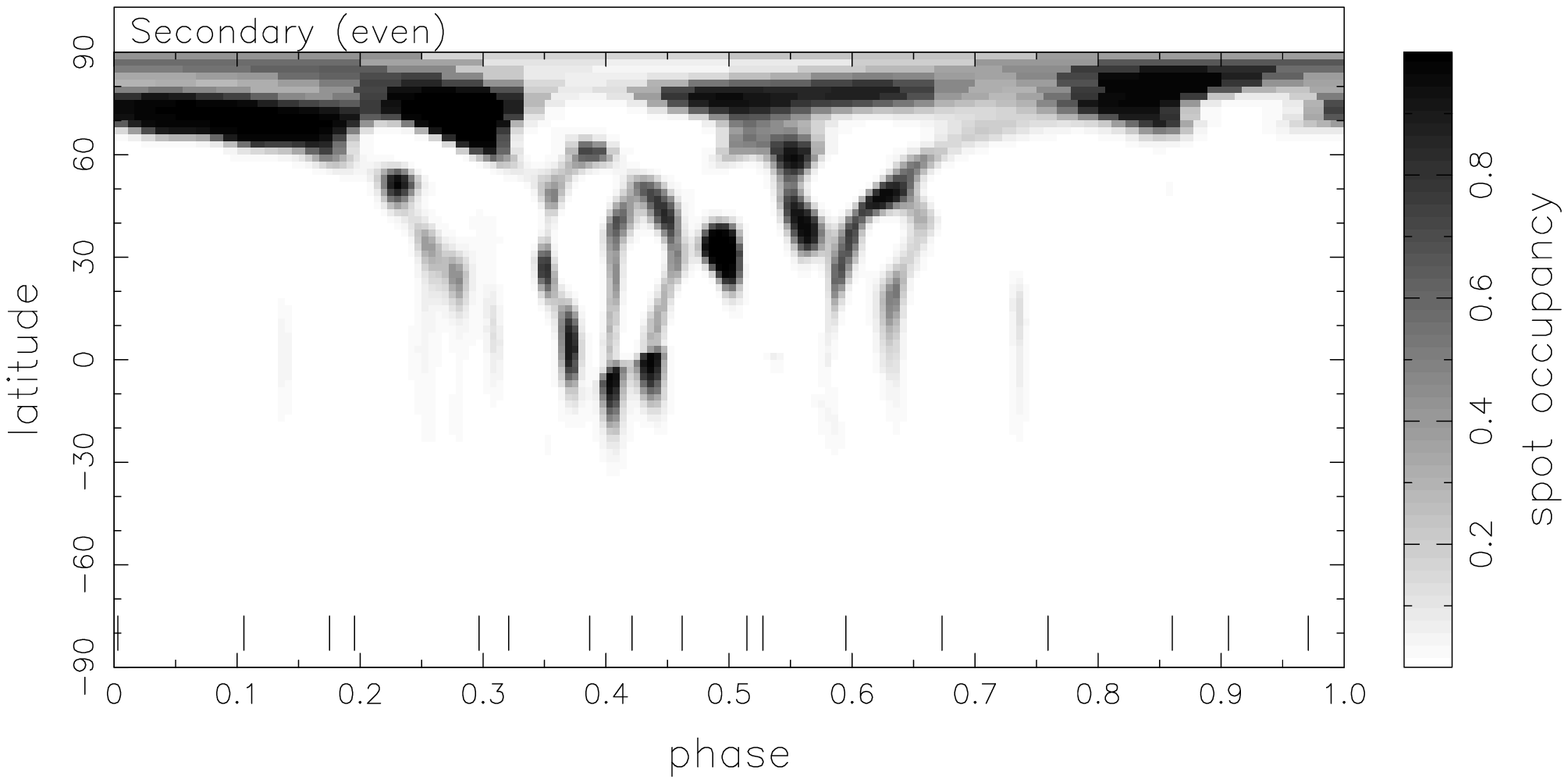}
\\
\end{tabular}
}

\subfloat[]{
\label{fig:test:phase}
\begin{tabular}{cc}
\includegraphics[angle=270,width=0.45\textwidth]{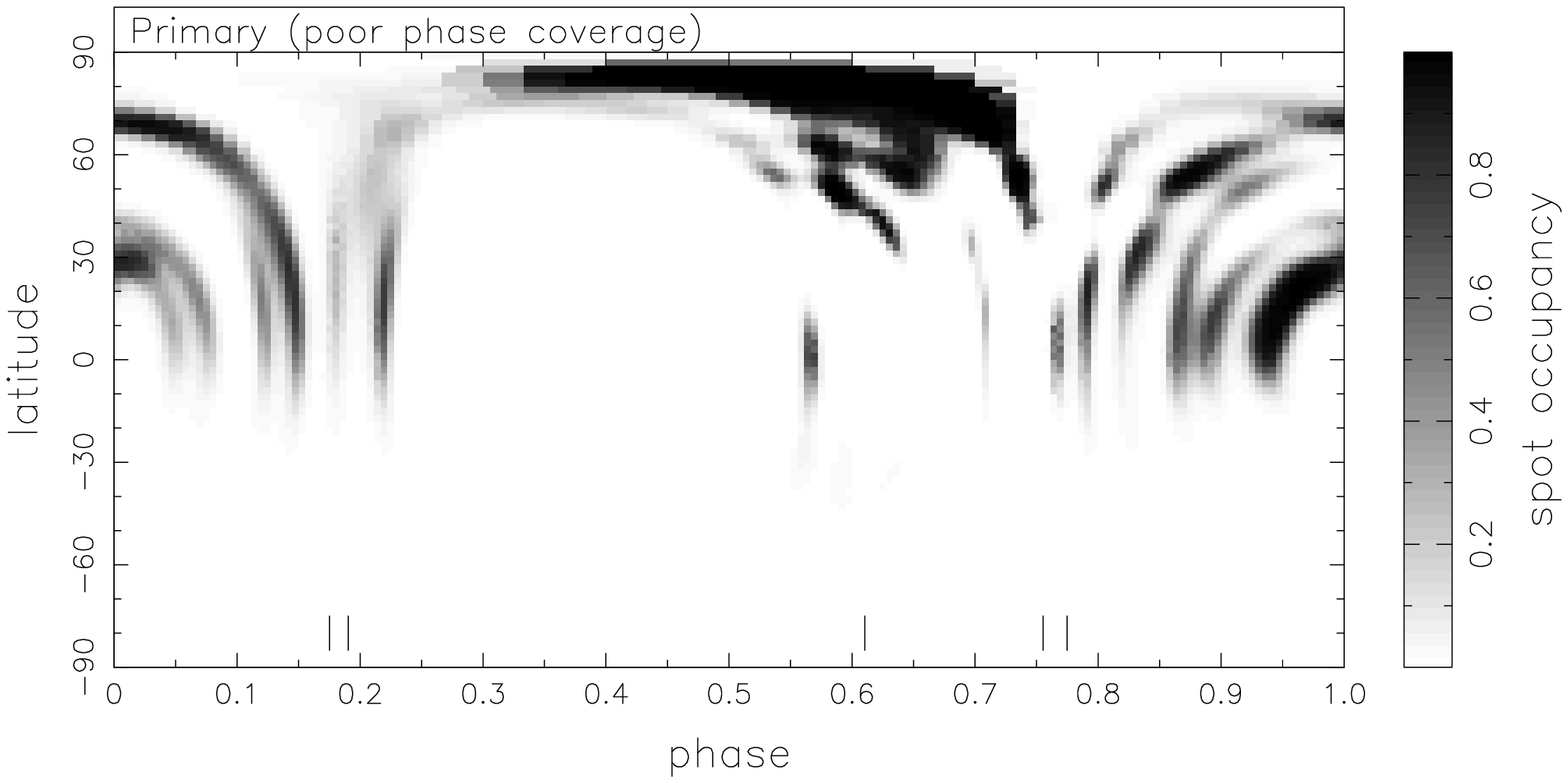}
&
\includegraphics[angle=270,width=0.45\textwidth]{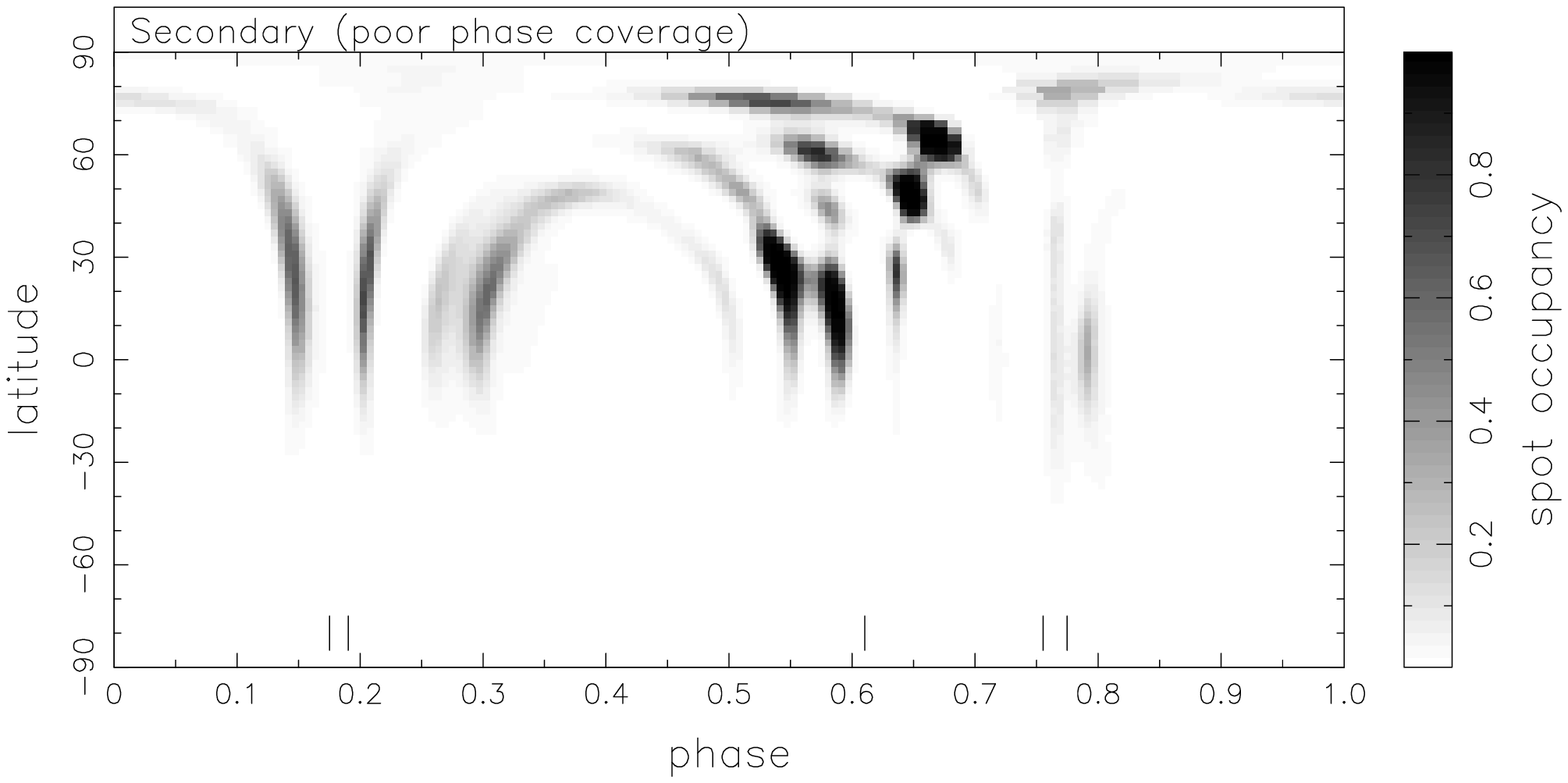}
\\
\end{tabular}
}

\caption{The surface images of 2008 derived from all (a), odd-numbered (b), even-numbered (c) and only five spectra (d).}
\label{fig:test}
\end{figure*}

In order to demonstrate the reliability of our image reconstruction, we made a test with the dataset obtained in 2008 November. The process is similar to the one in \citet{bar2004}. The dataset was divided into two series which were odd-numbered and even-numbered spectra. As a result, each series had 17 LSD profiles. Then we reconstructed surface images from these two data sub-sets, by using the same system parameters. The resulting surface maps of the primary and the secondary components are plotted in Fig. \ref{fig:test}. We also plot the images from all spectra in Fig. \ref{fig:test} for comparison. From the test images, we can see that the main starspot structures are quite similar, but some spots have slightly different strength and position. The differences between odd and even images are partly due to the S/N variation in the spectra. Because the code allocates the weighting factor to each profile according to its S/N, so the neighbouring odd and even profile pairs with different S/N make a different contribution to the individual image during the fitting procedure, which results in the differences between the odd and even images. In addition, such differences can also partly arise from the limitation of the image reconstruction with the current data, and some fine structures may be spurious.

Since the 2004 observations had much poorer phase sampling, we also performed a test to estimate the effect of the insufficient phase coverage on the reconstructed images of 2004. In the test, we chose five spectra from the 2008 dataset for the image reconstruction, which had similar phase values to the 2004 dataset. The resulting images are displayed in Fig. \ref{fig:test:phase}. The results show the smearing or absence of starspot features in the reconstruction. Although the large high-latitude spot around phase 0.6 on the primary star exists in the image, other high-latitude and polar spots on both components are much weaker or absent due to the large phase gaps in the data. In summary, due to their poor phase coverage, the 2004 images only contain information about the large-scale longitude distribution of spots on the primary. They do not reliably sample the spot distribution on the secondary. Furthermore, they do not tell us whether a polar spot existed on any of the components in 2004.

\section{Discussion}

\begin{figure*}
\centering
\begin{tabular}{ccc}
\centering
\subfloat[]{
\begin{minipage}[b]{0.3\textwidth}
\centering
\includegraphics[angle=270,width=0.97\textwidth]{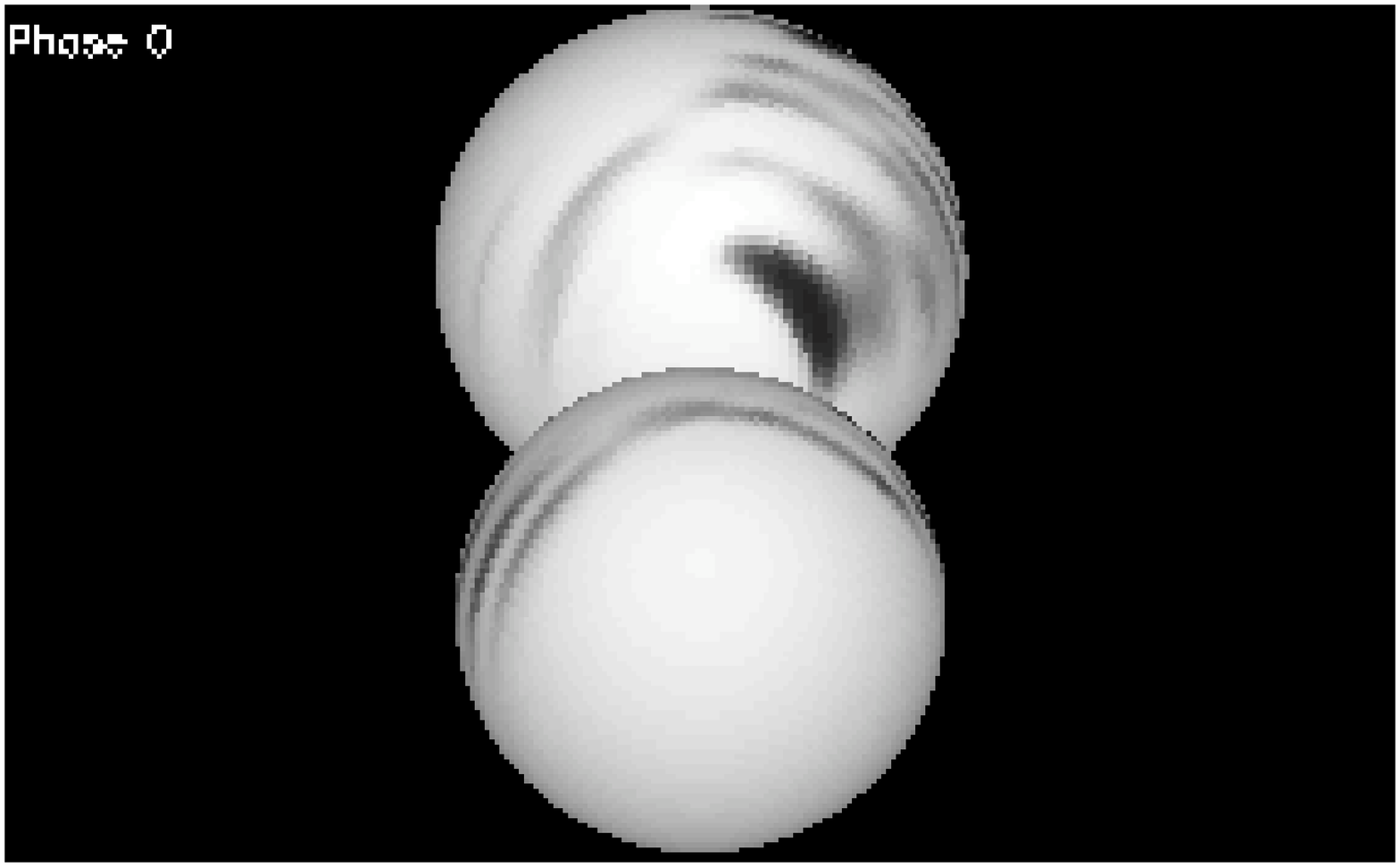}
\includegraphics[angle=270,width=0.97\textwidth]{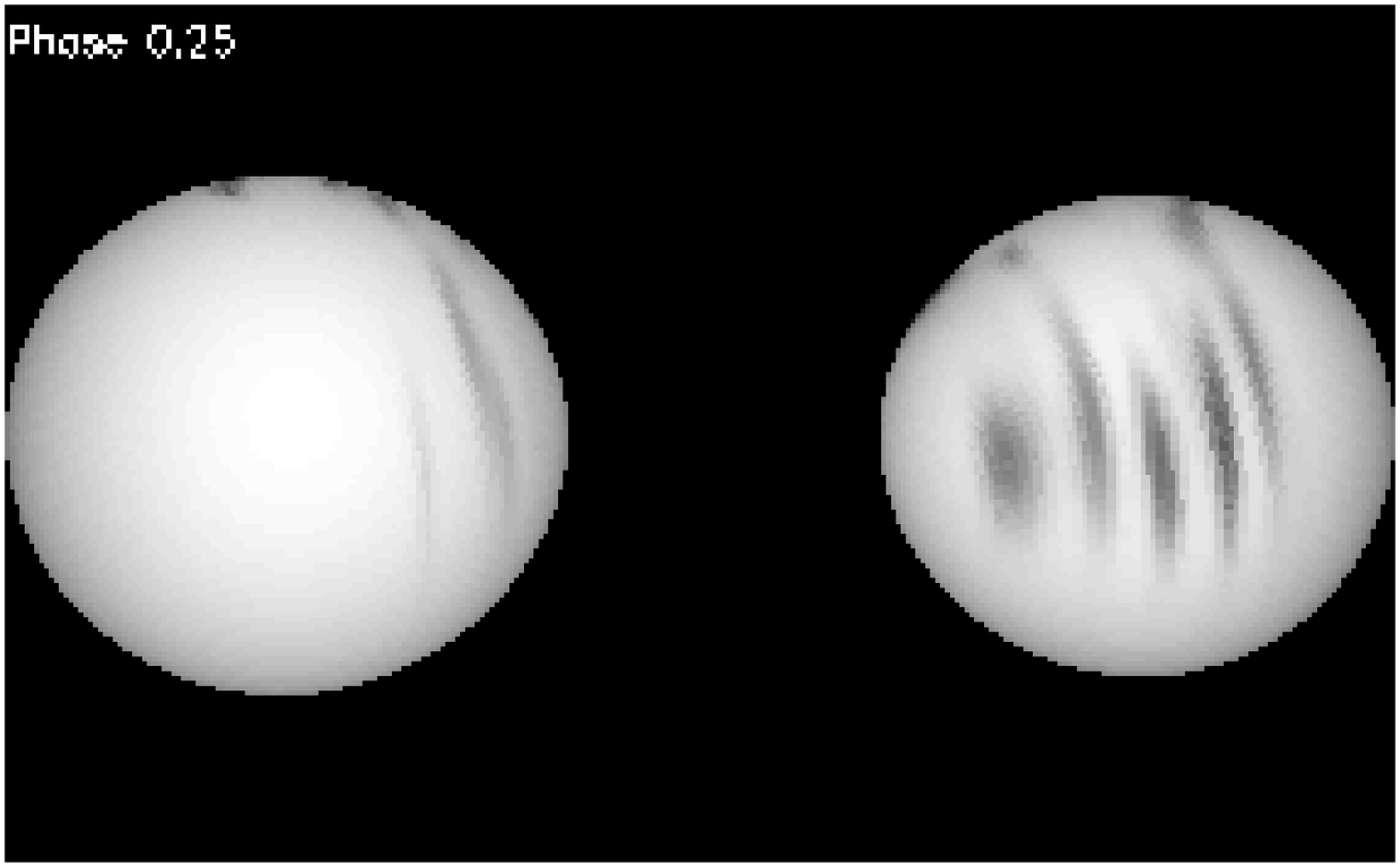}
\includegraphics[angle=270,width=0.97\textwidth]{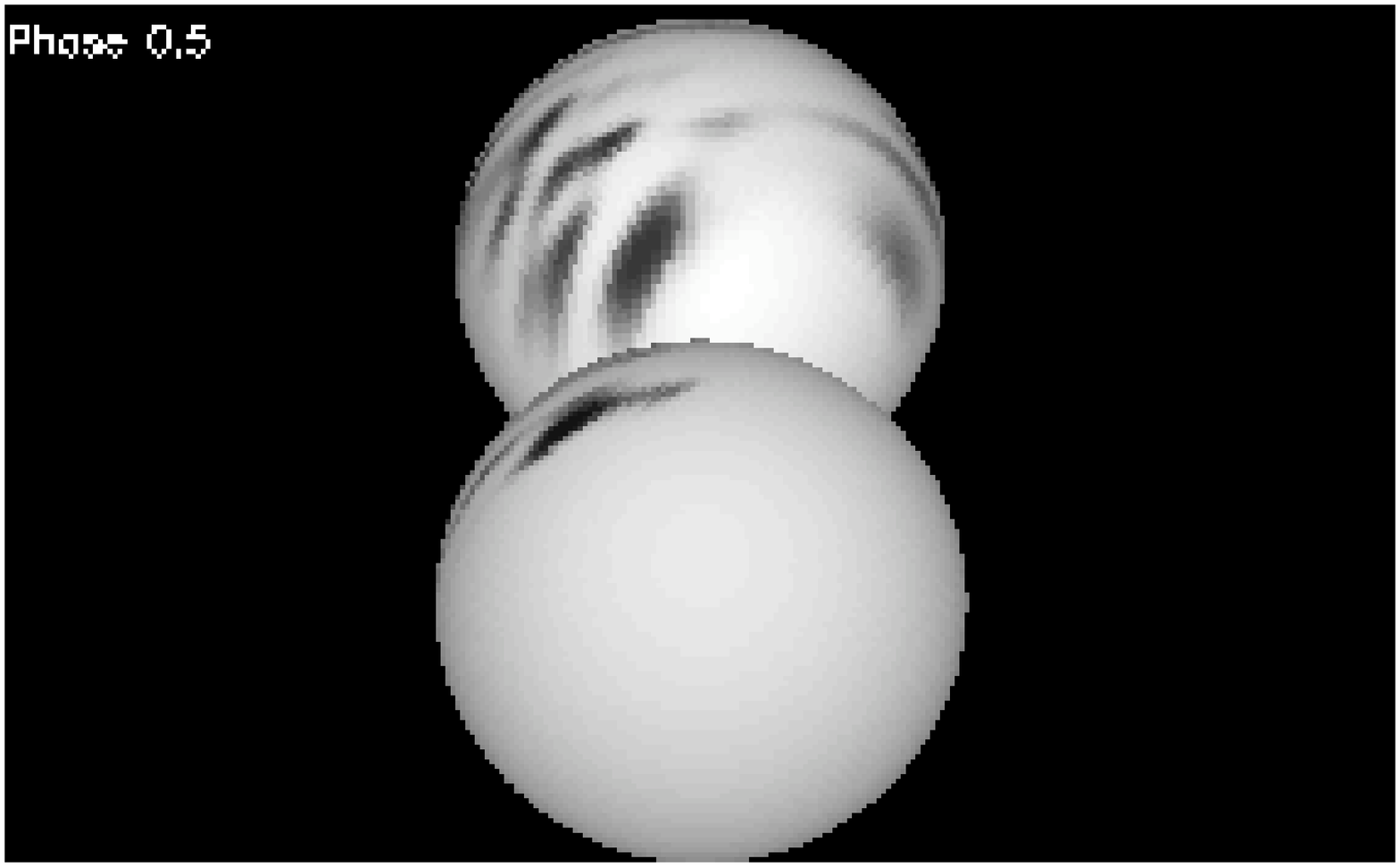}
\includegraphics[angle=270,width=0.97\textwidth]{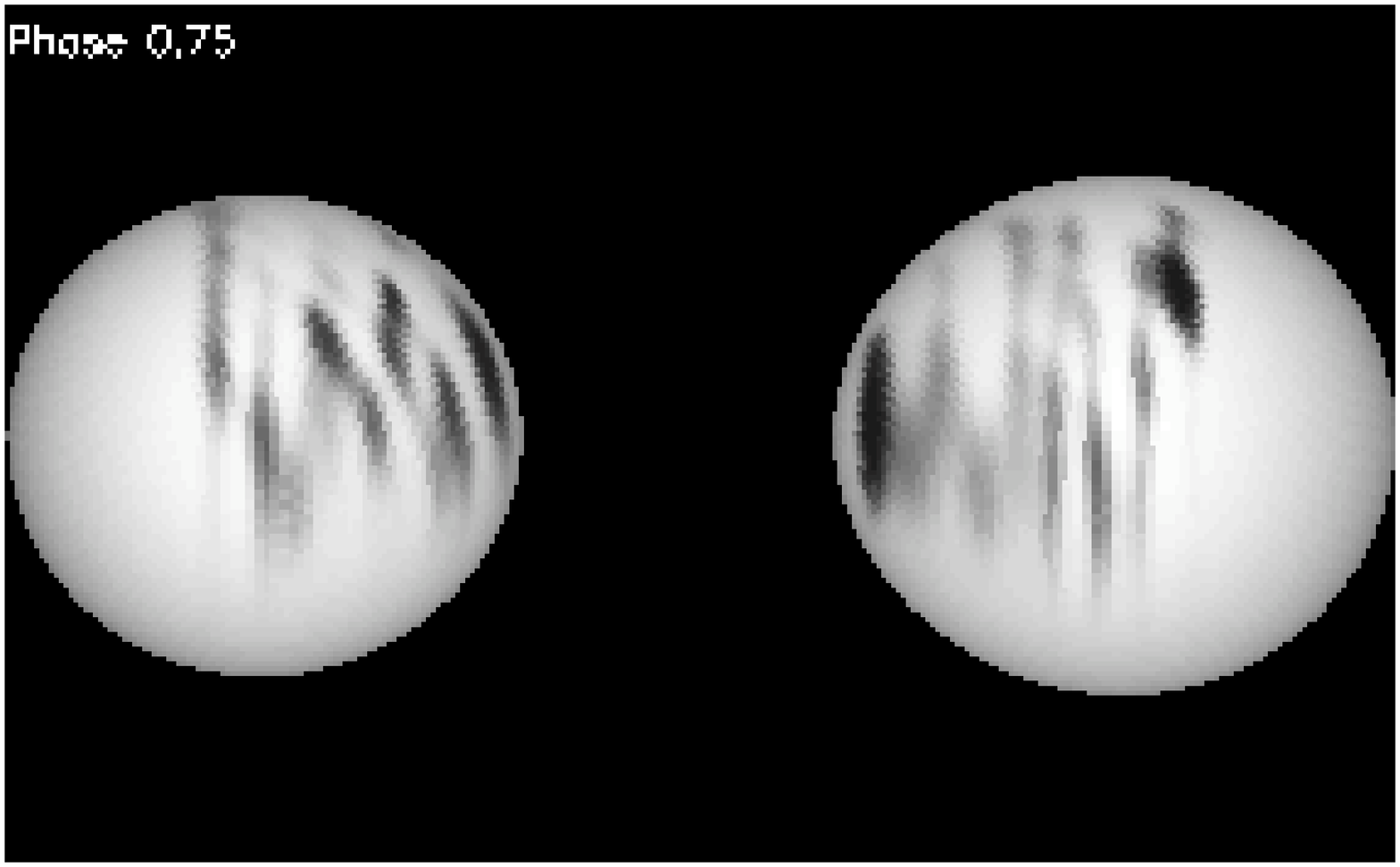}
\end{minipage}
}
&
\subfloat[]{
\begin{minipage}[b]{0.3\textwidth}
\centering
\includegraphics[angle=270,width=0.97\textwidth]{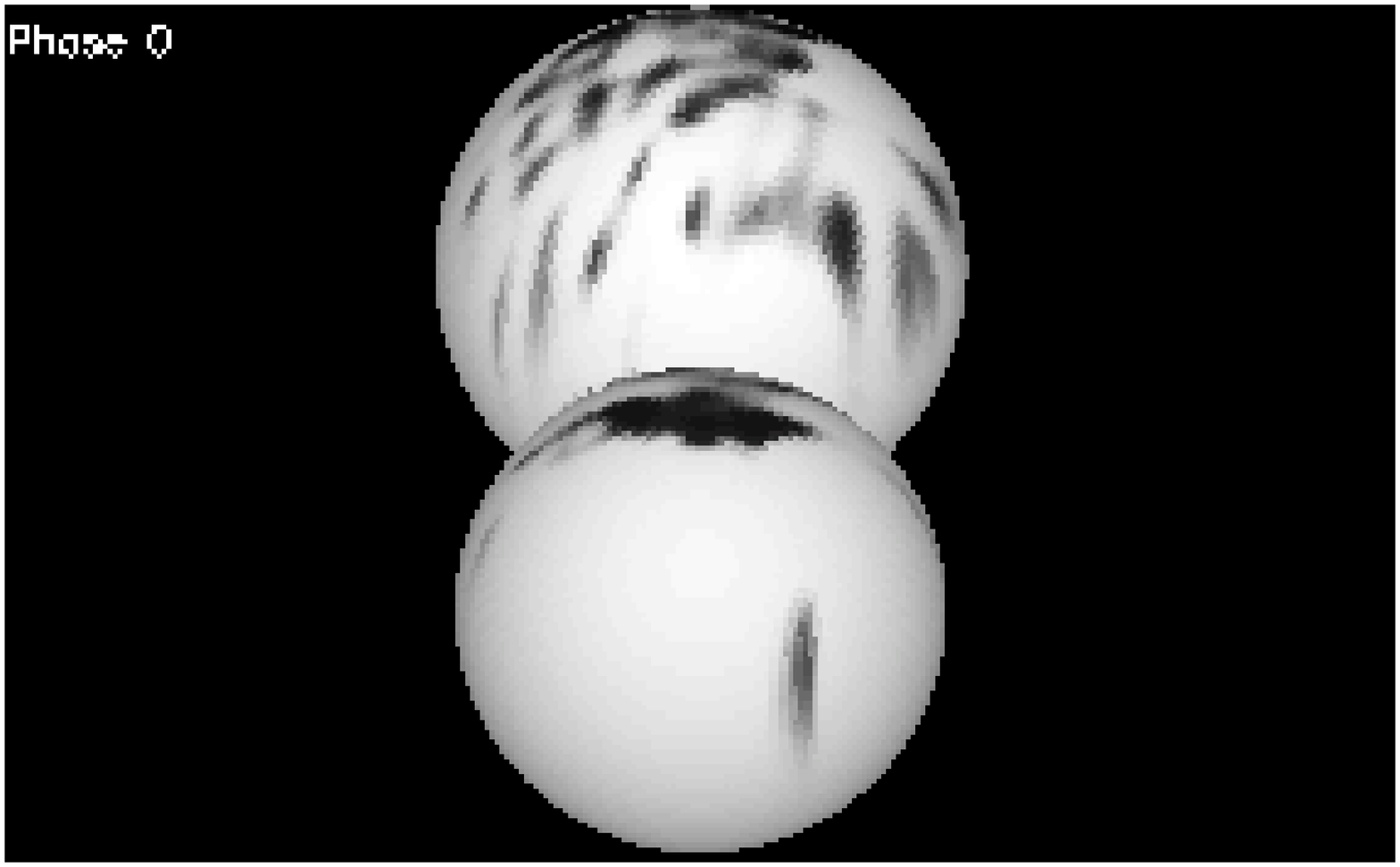}
\includegraphics[angle=270,width=0.97\textwidth]{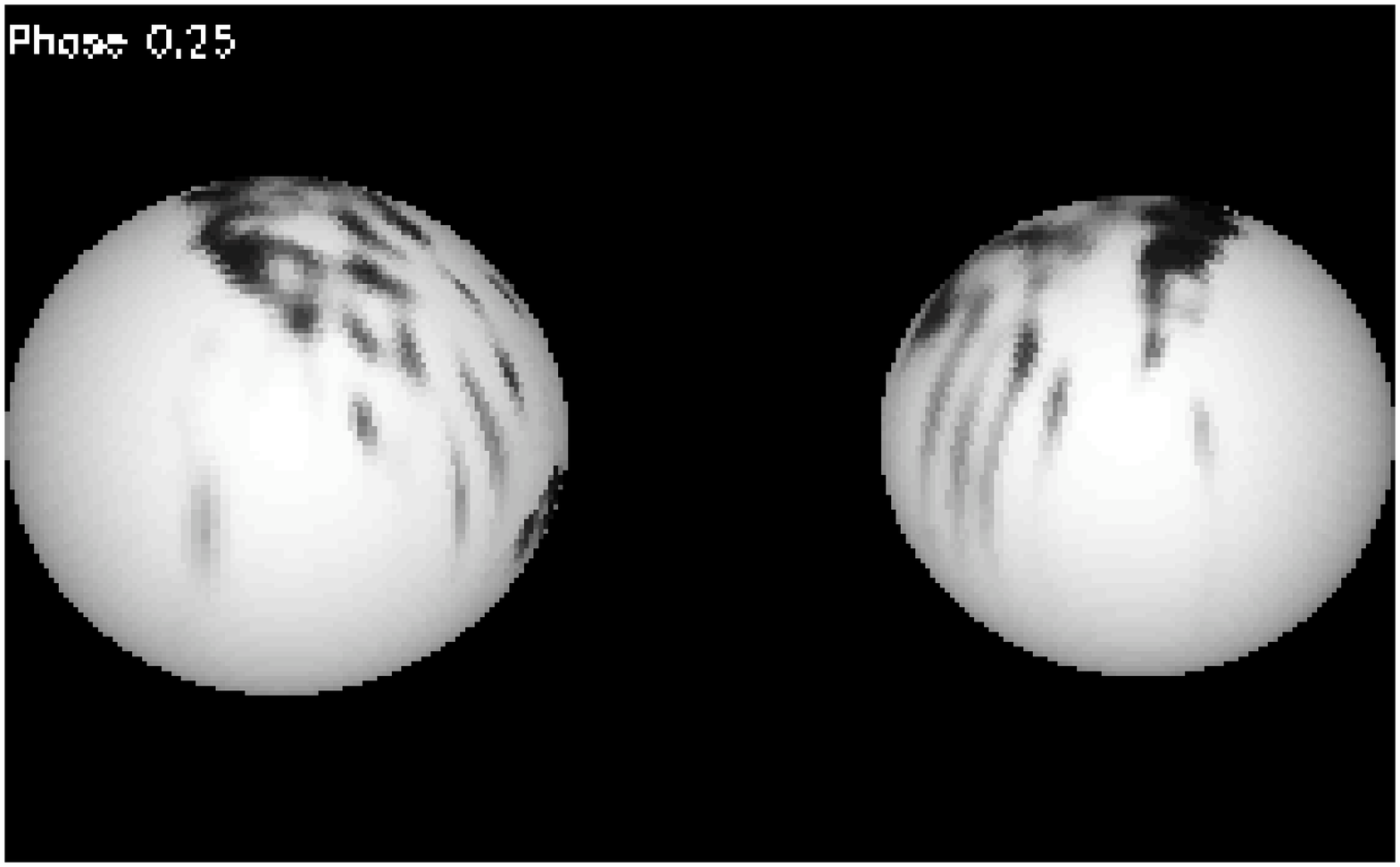}
\includegraphics[angle=270,width=0.97\textwidth]{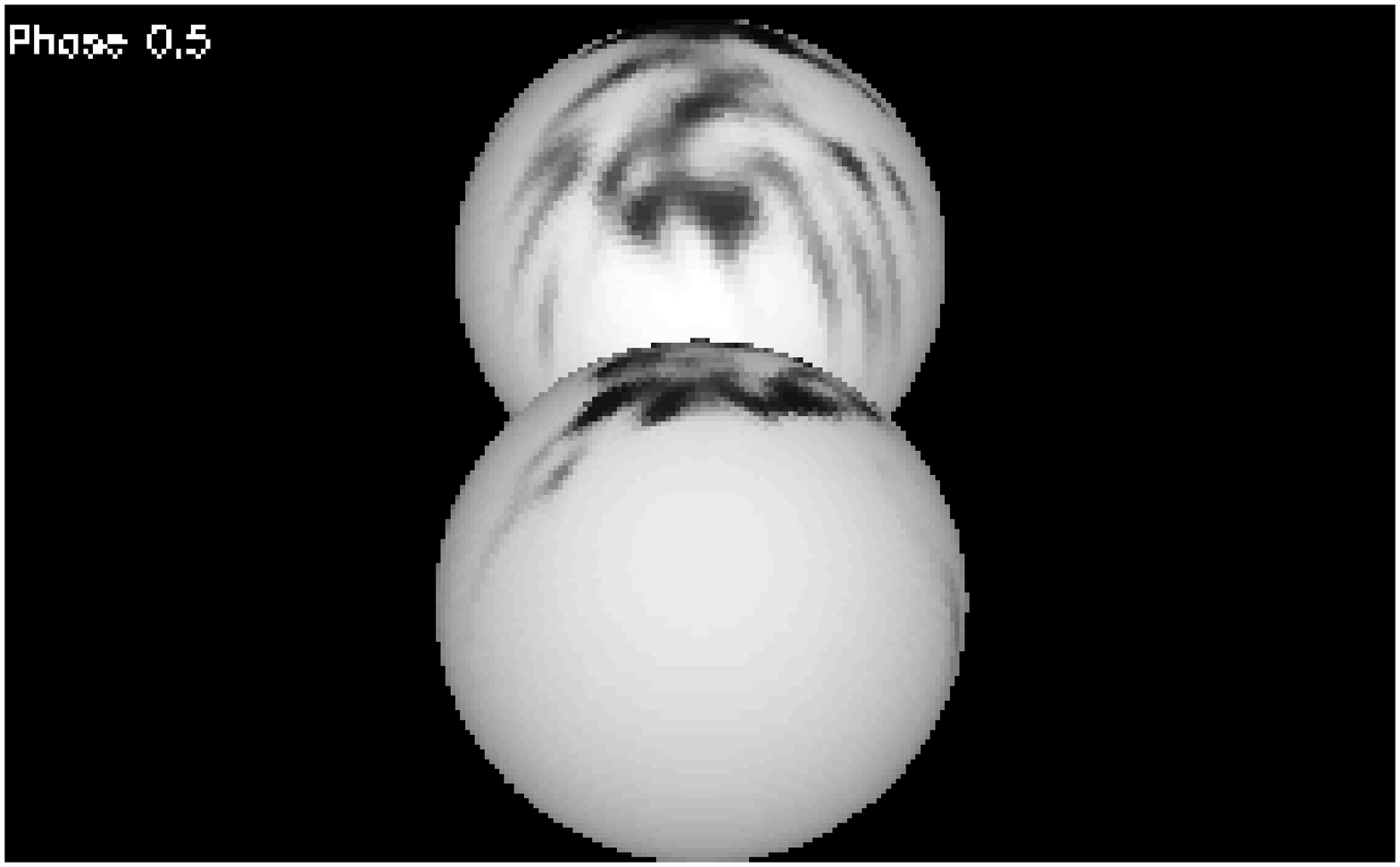}
\includegraphics[angle=270,width=0.97\textwidth]{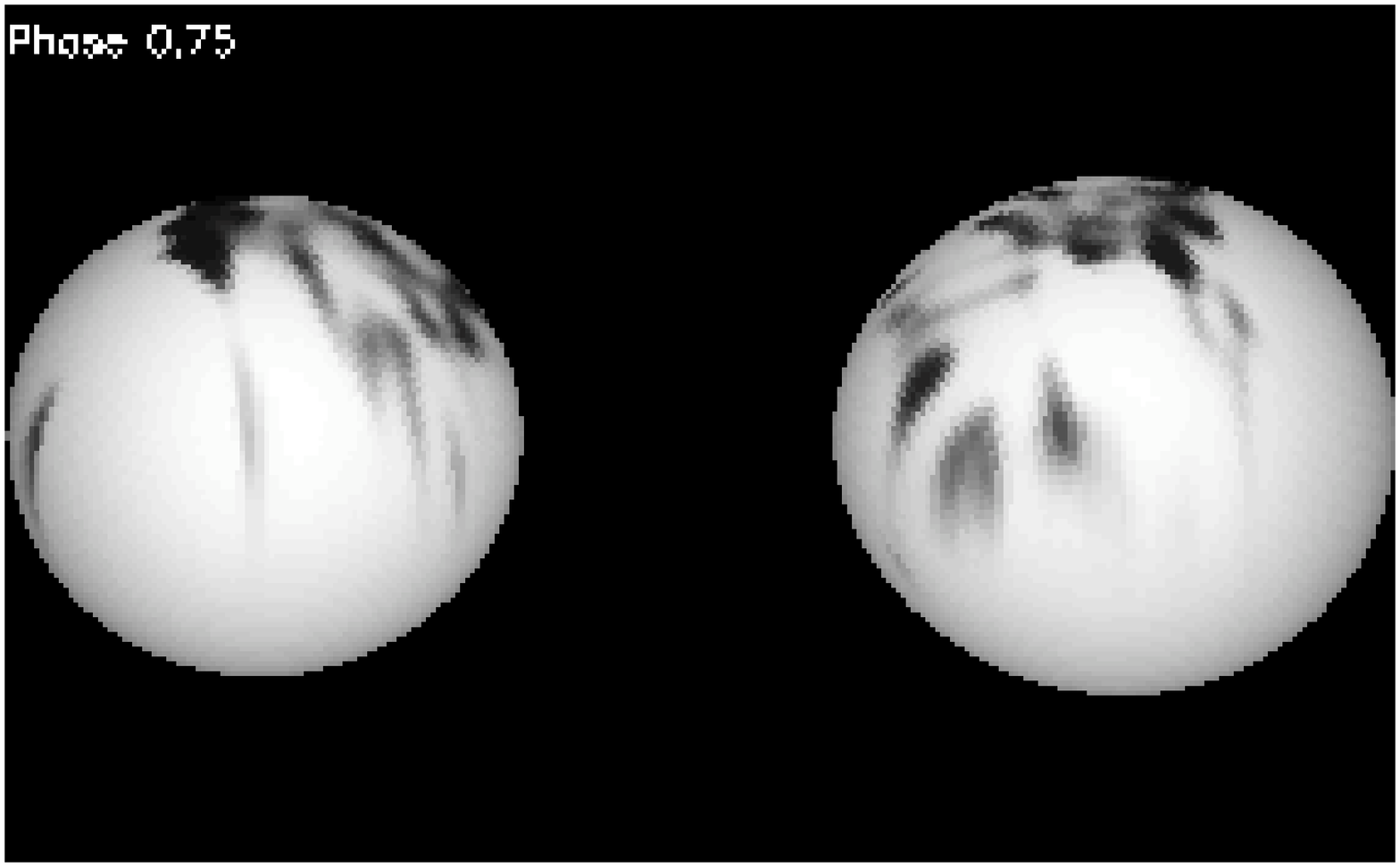}
\end{minipage}
}
&
\subfloat[]{
\begin{minipage}[b]{0.3\textwidth}
\centering
\includegraphics[angle=270,width=0.97\textwidth]{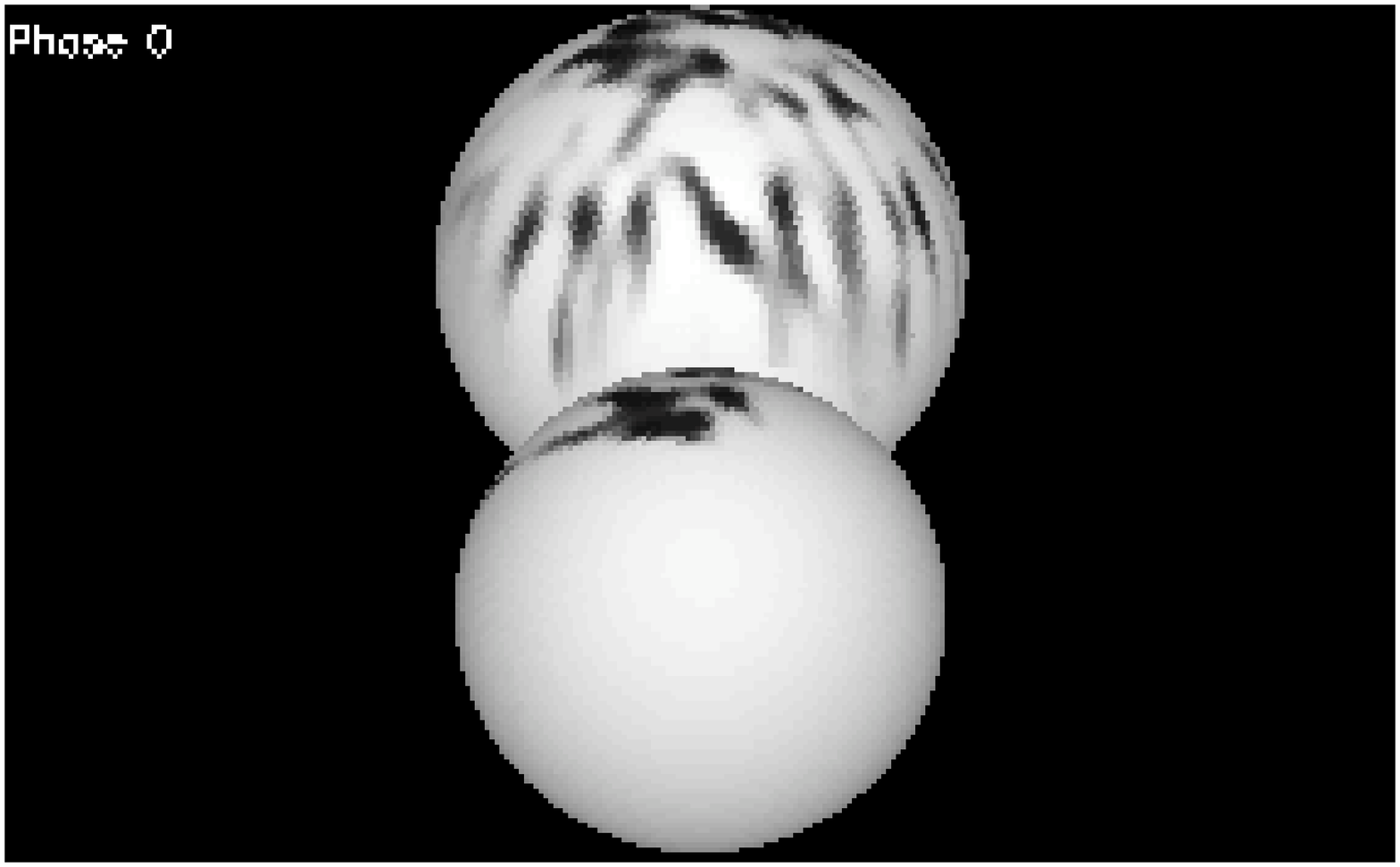}
\includegraphics[angle=270,width=0.97\textwidth]{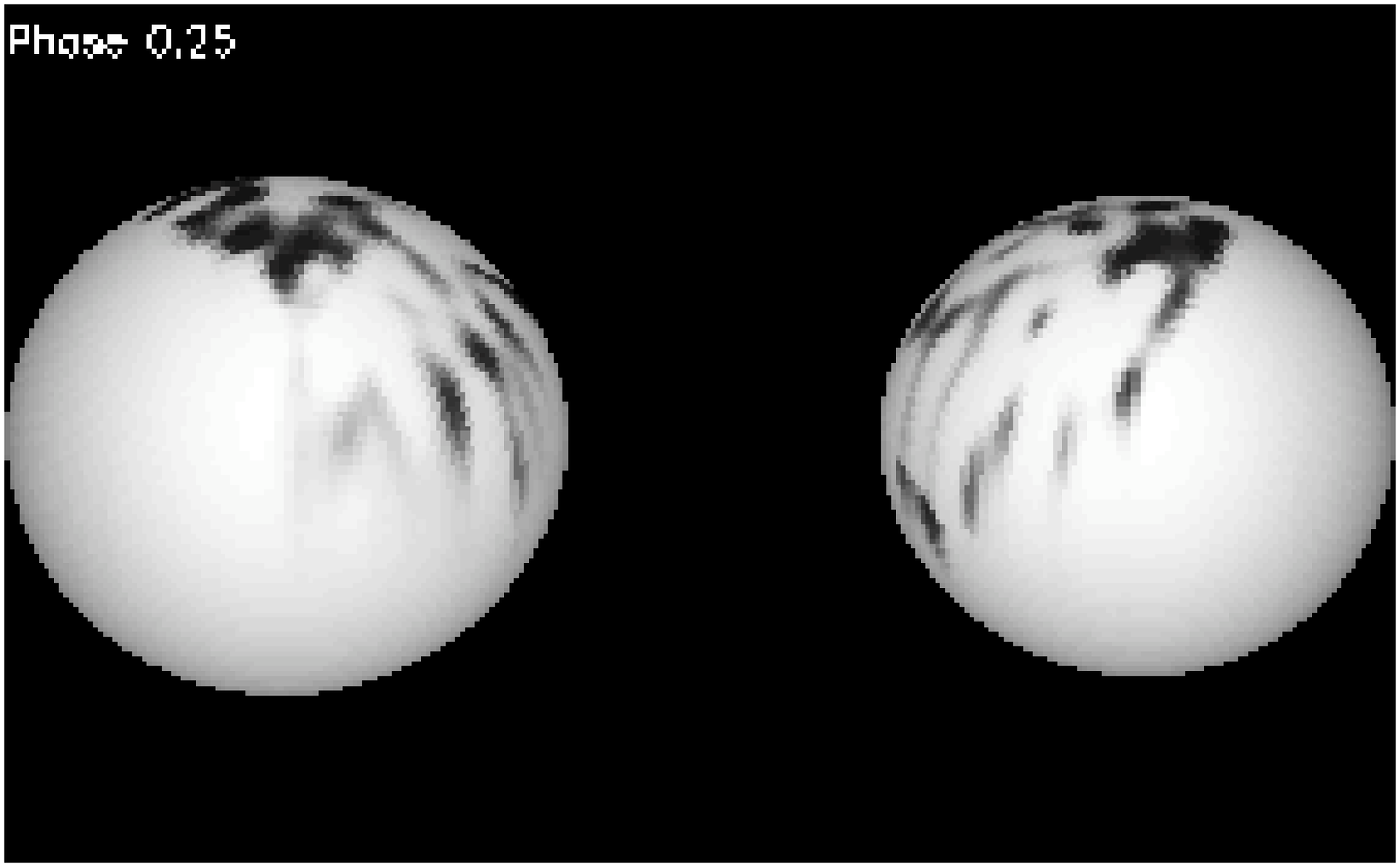}
\includegraphics[angle=270,width=0.97\textwidth]{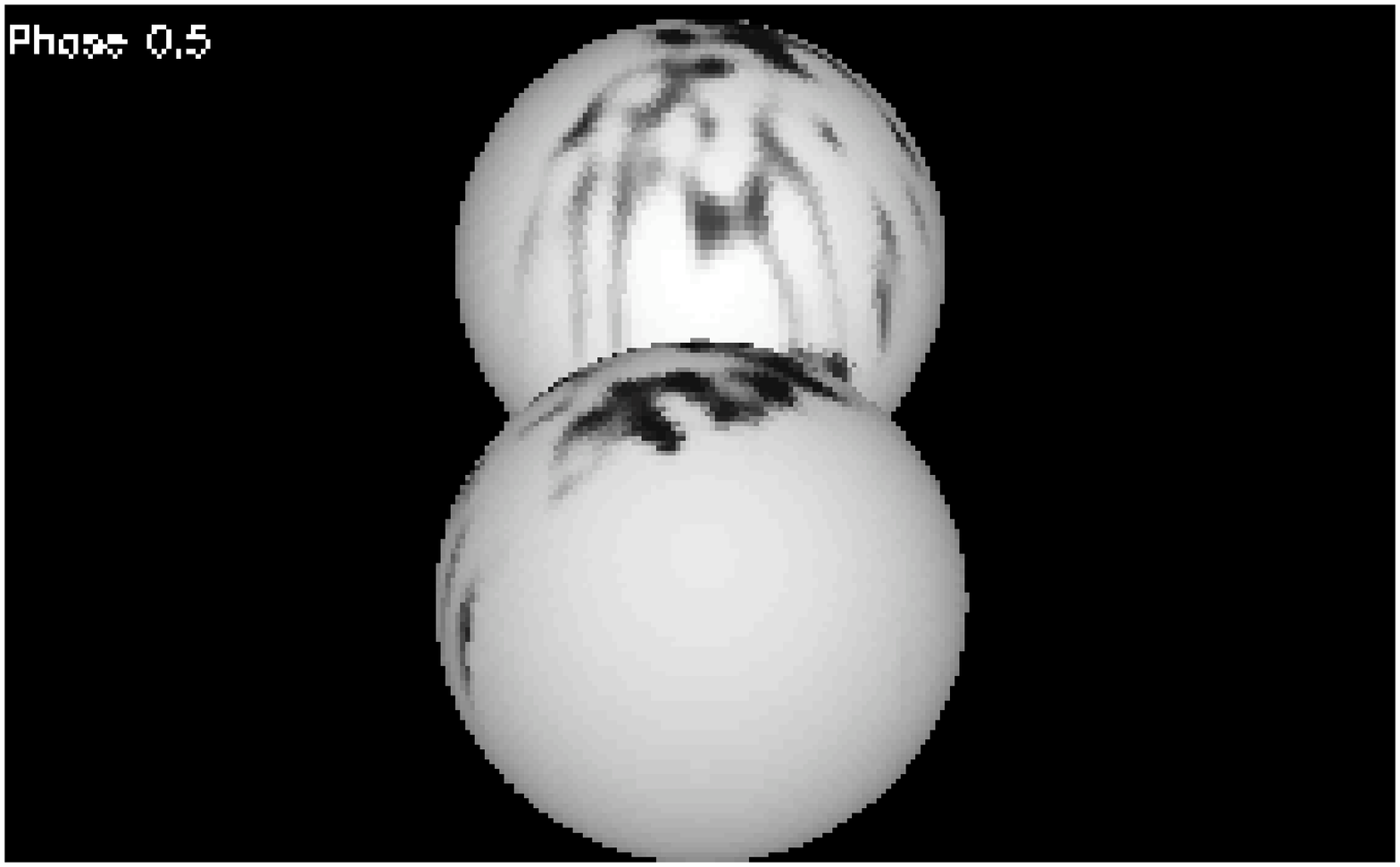}
\includegraphics[angle=270,width=0.97\textwidth]{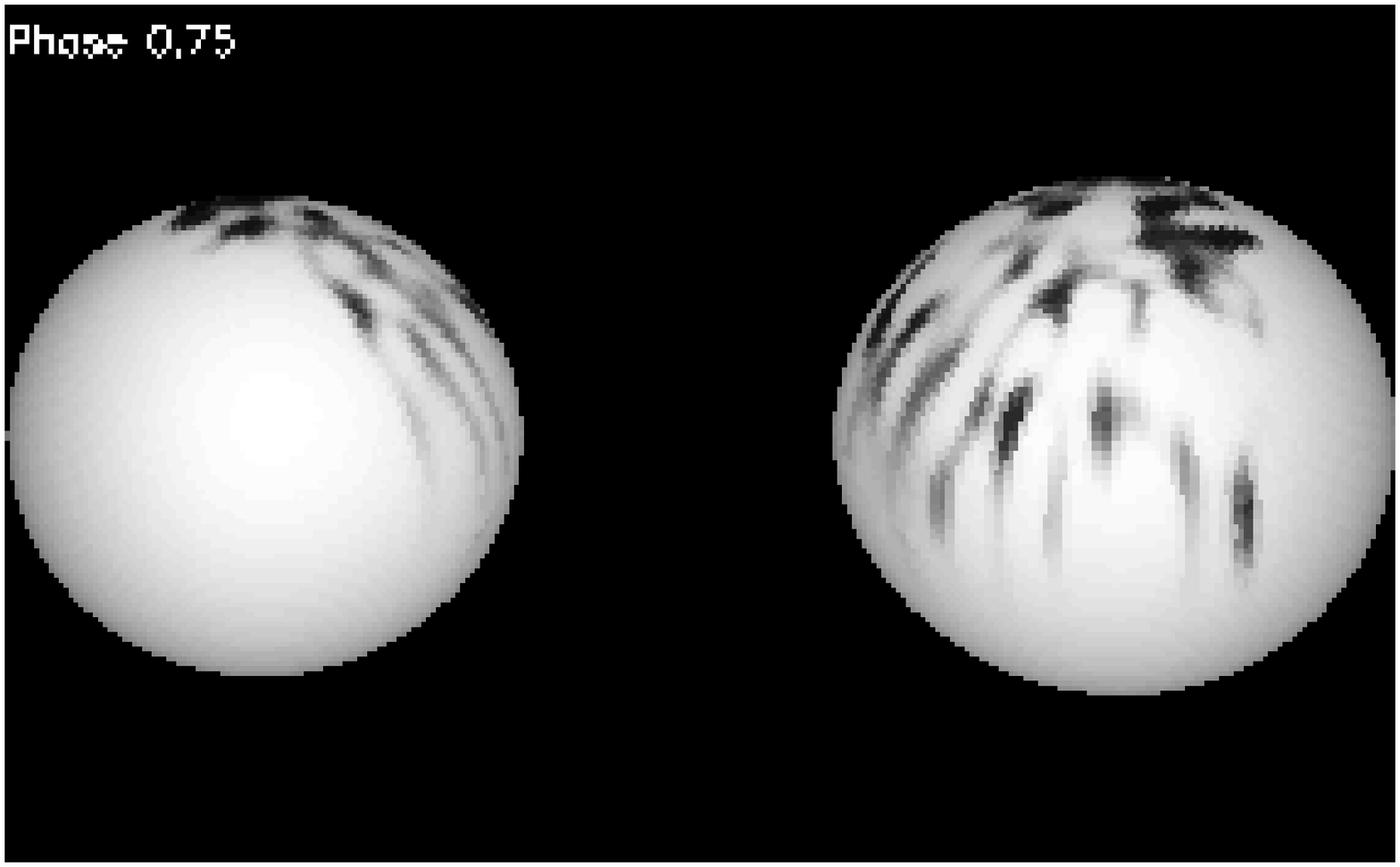}
\end{minipage}
}
\\
\end{tabular}
\caption{The images of the binary system ER Vul at orbital phases 0, 0.25, 0.5 and 0.75 for (a) 2004 November, (b) 2006 September and (c) 2008 November.}
\label{fig:images}
\end{figure*}

As can be seen in Fig. \ref{fig:primary} and Fig. \ref{fig:secondary}, both components of ER Vul exhibit strong starspot activities. The starspots can emerge at all latitudes on both stars during our observations. In the surface images for the primary and the secondary components of 2006 September and 2008 November, there were several extended starspots at the high-latitude belts (60\de--90\de). On the secondary star, the high-latitude starspots were much larger and can even reach the north pole of the star. Besides, there were also several starspots in the low and intermediate latitude regions. Some of them were attached to the high-latitude starspots. \citet{vin1993} had demonstrated that the eclipse information is useful for reducing the equatorial symmetry in Doppler images, based on numerical simulations for imaging eclipsing binary. So the low-latitude spots in our images are also reliable. The Doppler images derived by \citet{sok2002} also revealed starspots at various latitudes on both stars of ER Vul.

Both components of ER Vul are Sun-like stars, but have much higher rotational velocities (about 40$\Omega_{\odot}$) owing to the short period and the tidal lock effect. Different from the Sun, which only has sunspots at low latitudes, each component of ER Vul shows starspots at very high latitudes, even on the pole of the star. The size and strength of starspots found on ER Vul are also much larger than that of sunspots. The high-latitude or polar starspots on the surface of the rapidly rotating main-sequence stars are predicted by the numerical simulations of the magnetic flux tube models \citep{sch1996,gra2000}. Their models showed that the Coriolis force can be enhanced by the high rotational velocity, and push the flux tube arose from the bottom of the convection zone to the high-latitude regions. As a result, starspots can emerge at high latitudes or the poles of stars. The latitude of a starspot is related to the initial latitude of the flux tube and the rotational velocity of the star. \citet{sch2001} had shown that the magnetic flux can also be transported to the polar regions by the meridional advection on the active Sun-like stars, which can result in a strong polar cap magnetic field and polar starspots. \citet*{isik2011} had proposed a new magnetic flux model to simulate the flux generation and the transport in convention zone and on surface. Their simulations revealed the coexistence of low-to-mid latitude and strong polar starspot activities on the rapidly rotating Sun-like main-sequence stars.

From the Doppler images of 2006 September and 2008 November, which have good phase coverage, we can see that several pronounced high-latitude starspots on both stars existed at similar positions in two seasons, although the shape and strength changed. \citet{sok2002} had revealed the presence of large polar or near polar spots on both stars in 1994 and 1996. \citet{shk2005} also found that the large high-latitude active regions on one or both components of ER Vul were required to explain the observed Ca~{\sc ii} emission in 2002 August. The near polar regions on both components of ER Vul seem to exhibit a high level of activity in these seasons. \citet{hus2002} analysed the results of Doppler imaging and photometry studies on several single and binary stars and revealed that starspots on close binaries have much longer lifetimes than the ones on single stars. Some RS CVn-type binary stars have polar starspots for decades. However, the interval among our three observing runs, about two years, is too long to reveal the short-term evolution of the starspots on ER Vul. Besides, long-lived starspots are tracers for the active longitude migrations, which are common phenomena on RS CVn-type binary stars \citep{ber1998}. However, we find no clear evidence for the phase shift of the starspots on both components of ER Vul from our Doppler images of different seasons. The active regions on both stars seem to have fixed positions during our observations.

\begin{figure}
\centering
\includegraphics[angle=270,width=0.4\textwidth]{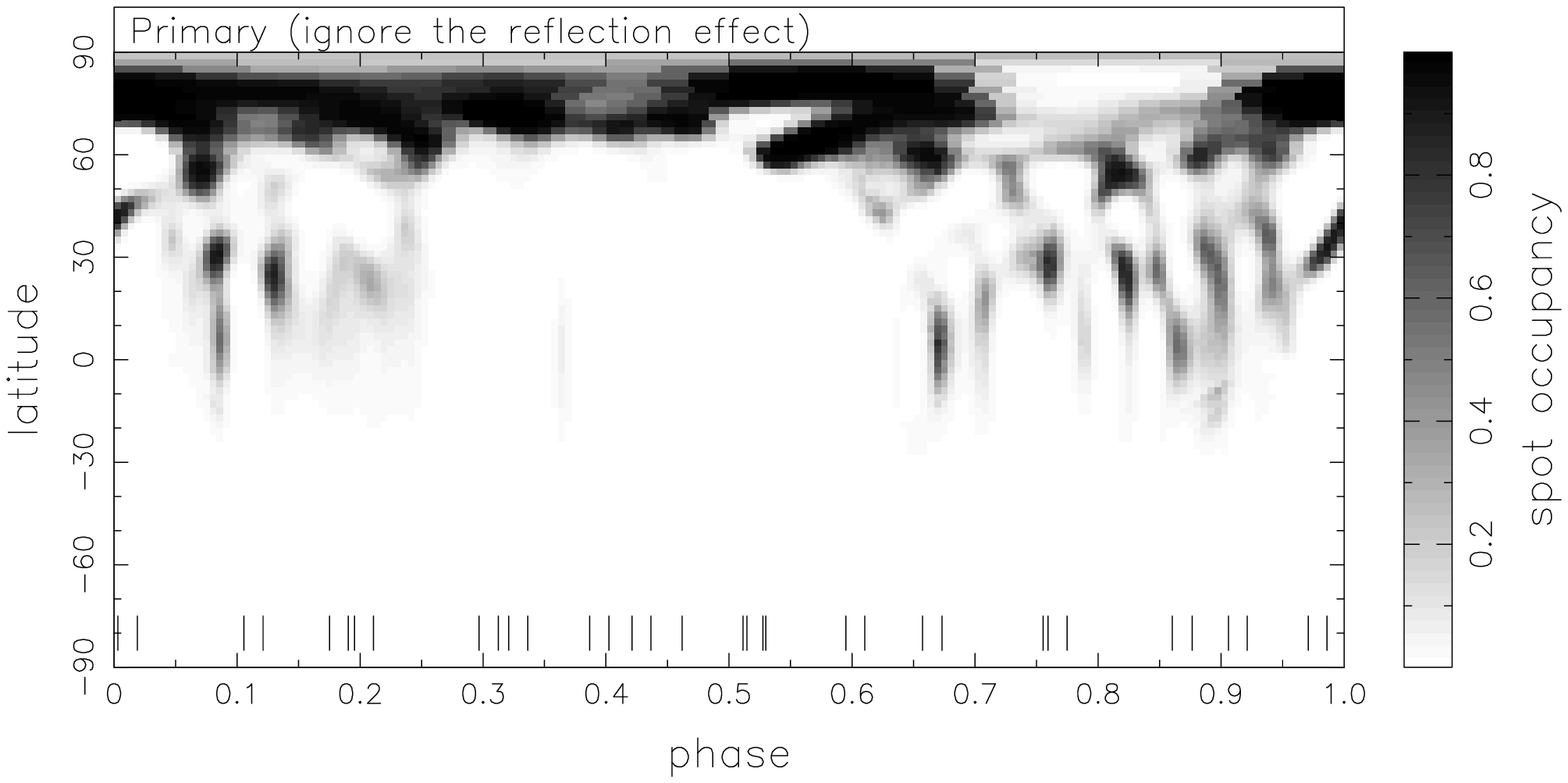}
\includegraphics[angle=270,width=0.4\textwidth]{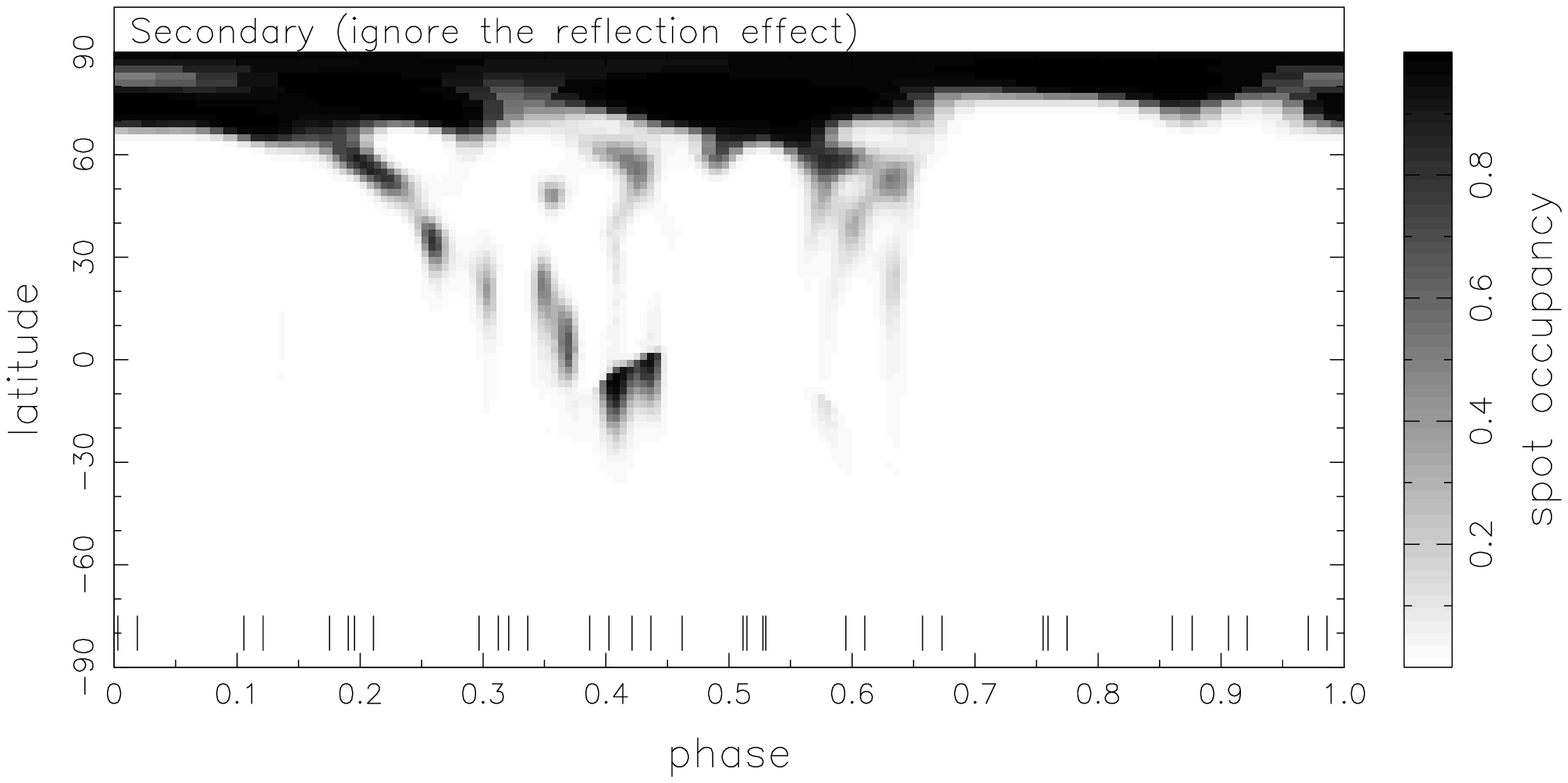}
\caption{The surface images derived from the 2008 dataset, but we forced the code to ignore the reflection effect in the computation. The high-latitude regions are more heavily spotted due to the errors in the equivalent width of the profiles introduced by the reflected light.}
\label{fig:albedo}
\end{figure}

The surface images reveal that the longitude distributions of starspots are not uniform on both components during our observations. The low-to-mid latitude spots were always concentrated at phase 0.7--1.3 on the primary star, and phase 0.3--0.7 on the secondary star. In order to show the relationship between starspot distribution and the relative positions of two component stars of ER Vul clearly, we plot the images of the binary system at four orbital phases (0, 0.25, 0.5 and 0.75) in Fig. \ref{fig:images}. As can be seen in the figure, most low- and intermediate-latitude starspots emerged on the hemisphere facing to the other star in ER Vul during our observations. The Doppler images derived by \citet{pis1996}, \citet{pis2001} and \citet{pis2008} also revealed non-uniform longitudinal spot distributions for ER Vul. In their spot maps, there were hot spots at the sub-stellar points and intermediate-latitude cool starspots at fixed longitudes. Our image code cannot reconstruct bright photospheric features because it uses the two-temperature model \citep{cam1992}. However, the presence of a hot spot without the absorption feature can produce the same line deformation as a completely dark cool spot, but change the equivalent width of the profile obviously \citep*{unr1998}. According to \citet{pis2008}, the bright features were produced by the reflection effect, and mainly concentrated at the sub-stellar point of each star. If we ignore the reflected light in the reconstruction, high-latitude regions will be much more heavily spotted due to the errors in the equivalent width of the profiles (Fig. \ref{fig:albedo}). In our case, since the reflection effect had been considered in the computation \citep{cam1997}, we think the influence on our reconstructed images was eliminated. The low-to-mid latitude starspot structures in our images are similar to the cool starspot patterns in their Doppler images. Besides, \citet{shk2005} also found a active region near the sub-binary longitude on the secondary star of ER Vul from light curve modelling. We may infer that the non-uniform longitude distribution of starspots on ER Vul is persistent.

The similar non-uniform longitudinal spot distributions on other close binary systems had also been reported. In the recent Doppler images of the pre-main-sequence binary V824 Ara (G5IV + K0IV), \cite{kri2013} found hot spots on both components facing to each other. They suggested that these features indicated the magnetic interaction in the binary. \citet{str2003} imaged the close binary system $\sigma^{2}$ CrB (F9V + G0V), which is similar to ER Vul, and found the primary star had cool starspots facing to the secondary star. They also found equatorial warm belts at the trailing hemispheres of both stars of $\sigma^{2}$ CrB. \citet{gar2003} also revealed mirror starspots on the stars of the RS CVn-type binary HR 1099.

\citet{hol2003} had shown that the tidal force can influence the arising flux tube and lead to the formation of the preferred longitudes, which are usually found on RS CVn-type binary stars. From the simulations of stellar dynamo models with the parameters of ER Vul, \citet*{moss2002} demonstrated that the tidal interaction can cause a non-uniform longitude distribution of the magnetic structure on the close binary systems like ER Vul. The differential rotations cannot smooth the non-uniform distribution on the tidal locked binary stars. On the other hand, the magnetic interaction is also common in close binary systems. \citet{uch1985} had proposed a model for the interacting magnetospheres of the RS CVn-type binaries. They revealed that two component stars in the close binaries are magnetic connecting, and the active longitude belts are related to the interaction between magnetic fields. \citet{pri2000} found starspots on both stars of the short-period RS CVn-type binary RT And were facing to each other. They inferred that there was a magnetic bridge connecting two active regions on the component stars, which may lead to a mass transfer between two stars.

For close binaries, \citet{olah2006} suggested that the tidal force and the common magnetic field are the most important factors which affect the magnetic flux tube eruption and determine the starspot position. She summarized the observational results on several active close binary stars and found that the magnetic field interaction has more effects on the starspot activities of the main-sequence stars in the close binaries than the tidal force, because these stars have much higher surface gravity. In consequence, the main-sequence stars often show active regions at quadrature phases. ER Vul has two similar main-sequence stars with the surface gravity of log $g$ = 4.5, according to \citet{har2004}. However, the active regions can be located at the quadrature phases and near the sub-stellar points in our surface maps of ER Vul. Following the reasoning of \citet{olah2006}, this may indicate that the common magnetic field and the tidal effect are balanced for ER Vul. However in her samples, only subgiant stars show this mixed behaviour.

\section{Conclusion}
We have presented the Doppler images of both components of the short-period eclipsing binary ER Vul in 2004, 2006 and 2008 seasons. Based on the new surface maps, we summarize the results as follows.

1. From the new Doppler images, we can find starspots on both component stars of ER Vul. The starspots can be located at all latitudes on both stars of ER Vul.

2. Both components of ER Vul showed pronounced high-latitude starspots in 2006 and 2008. No obvious phase shifts of the starspots are found during our observations.

3. The longitude distributions of starspots are non-uniform on both stars of ER Vul. The hemisphere facing to the other component star is more active, especially in low-to-mid latitude regions.

In the future, more observations with shorter interval and longer time baseline are required to reveal the short-term evolution of starspots and the possible stellar activity cycle on both component stars of the close binary ER Vul.

\section*{Acknowledgements}
This work has made use of the VALD database, operated at Uppsala University, the Institute of Astronomy RAS in Moscow, and the University of Vienna. We would like to thank Profs. Jian-yan Wei and Xiao-jun Jiang for the allocation of observing time at Xinglong 2.16m telescope. We also thank the referee, Uwe Wolter, for helpful comments and suggestions that have significantly improved the clarity and quality of this paper. This work is supported by National Natural Science Foundation of China through grants Nos. 10373023, 10773027 and 11333006, Chinese Academy of Sciences through project No. KJCX2-YW-T24.

\end{document}